\documentclass[fleqn]{2023SCGE}
\pdfoutput=1
\DeclareUnicodeCharacter{2212}{-}
\usepackage[export]{adjustbox}
\begin{document}

\ensubject{subject}

\ArticleType{Article}
\SpecialTopic{SPECIAL TOPIC: }
\Year{2023}
\Month{xxx}
\Vol{xx}
\No{x}
\DOI{xx}
\ArtNo{000000}
\ReceiveDate{xxx xxx, 2023}
\AcceptDate{xxx xxx, 2023}

\title{Improving Constraint on $\Omega_{m}$ from SDSS Using Marked Correlation Functions}{Improving Constraint on $\Omega_{m}$ from SDSS Using Marked Correlation Functions}

\author[1,2]{Limin Lai}{}
\author[1]{Jiacheng Ding}{}
\author[1,2]{Xiaolin Luo}{}
\author[1,2]{Yizhao Yang}{}
\author[1]{Zihan Wang}{}
\author[1]{Keshi Liu}{}
\author[1,3]{Guanfu Liu}{}
\author[1,4]{\\Xin Wang}{}
\author[1,4]{Yi Zheng}{}
\author[2]{Zhaoyu Li}{}
\author[1,4,5]{Le Zhang}{zhangle7@mail.sysu.edu.cn}
\author[1,4]{Xiaodong Li}{lixiaod25@mail.sysu.edu.cn}

\AuthorMark{Lai L.M.}


\AuthorCitation{Lai L.M., Ding J.C., Luo X.L. et al.}

\address[1]{School of Physics and Astronomy, Sun Yat-sen University Zhuhai Campus, Zhuhai 519082, P. R. China}
\address[2]{Department of Astronomy, Shanghai Jiao Tong University, Shanghai 200240, P. R. China}
\address[3]{Department of Astronomy, Tsinghua University, Beijing 100084, P. R. China}
\address[4]{CSST Science Center for the Guangdong-Hong Kong-Macau Greater Bay Area, SYSU, Zhuhai 519082, P. R. China}
\address[5]{Peng Cheng Laboratory, No. 2, Xingke 1st Street, Shenzhen 518000, P. R. China}

\abstract{
Large-scale structure (LSS) surveys will increasingly provide stringent constraints on our cosmological models. Recently, the density-marked correlation function (MCF) has been introduced, offering an easily computable density-correlation statistic. Simulations have demonstrated that MCFs offer additional, independent constraints on cosmological models beyond the standard two-point correlation (2PCF). In this study, we apply MCFs for the first time to SDSS CMASS data, aiming to investigate the statistical information regarding clustering and anisotropy properties in the Universe and assess the performance of various weighting schemes in MCFs. Upon analyzing the CMASS data, we observe that, by combining different weights ($\alpha = [-0.2, 0, 0.2, 0.6]$), the MCFs provide a tight and independent constraint on the cosmological parameter $\Omega_m$, yielding $\Omega_m = 0.293 \pm0.006$ at the $1\sigma$ level, which represents a significant reduction in the statistical error by a factor of 3.4 compared to that from 2PCF. Our constraint is consistent with recent findings from the small-scale clustering of BOSS galaxies~\cite{zhai2022} within the 1$\sigma$ level. However, we also find that our estimate is lower than the Planck measurements by about 2.6$\sigma$, indicating the potential presence of new physics beyond the standard cosmological model if all the systematics are fully corrected. The method outlined in this study can be extended to other surveys and datasets, allowing for the constraint of other cosmological parameters. Additionally, it serves as a valuable tool for forthcoming emulator analysis on the Chinese Space Station Telescope (CSST).
}




\maketitle

\begin{multicols}{2}

\section{Introduction}\label{sec:intro}

The detection of cosmic acceleration~\cite{Riess_1998,Perlmutter_1999} suggests that our universe contains either a mysterious "dark energy" component or that our current understanding of general relativity might be incomplete on cosmological scales.  The pursuit of \Authorfootnote a theoretical explanation and the use of observational methods to study cosmic acceleration has gained significant attention. However, our understanding of this phenomenon and our ability to measure it accurately need further enhancement~\cite{Weinberg_1989, Li_2011, YOO_2012, Weinberg_2013}. At scales spanning hundreds of Megaparsecs (Mpc), galaxies are arranged into an intricate filamentary structure referred to as the "cosmic web"~\cite{Bardeen_1986, deLapparent, Huchra, Tegmark, Guzzo}. Within this cosmic web, the distribution and clustering characteristics of galaxies hold a wealth of information concerning the history of cosmic expansion and structure evolution.

In addition to optical surveys, 21 cm intensity mapping surveys will offer an independent mapping of the Universe, enabling constraints on the evolution of the Universe at higher redshifts~\cite{wu2023prospects,xu2020cosmological,zhang2023cosmology}. Moreover, space-based laser interferometers such as LISA~\cite{armano2016sub}, Taiji~\cite{hu2017taiji}, and TianQin~\cite{luo2020first} measure the luminosity distance of gravitational wave (GW) sources. The potential of GW standard sirens from the Taiji-TianQin-LISA network to constrain cosmological parameters is significant~\cite{wang2022forecast}, effectively breaking down degeneracies and providing independent measurements of the Hubble constant.

In the upcoming decade, several extensive surveys, such as EUCLID~\cite{EUCLID}, LSST~\cite{LSST}, WFIRST~\cite{WFIRST}, and CSST~\cite{Gong_2019}, will provide an immense volume of data for mapping the large-scale universe with unprecedented precision~\cite{li2023forecast}.  Consequently, it is crucial to develop robust statistical tools for a comprehensive and reliable analysis of cosmological models and the estimation of cosmological parameters from the cosmic large-scale structure (LSS).

In literature~\cite{Kaiser,Ballinger,Eisenstein_1998,
Blake_2003,Seo_2003}, measurements of the two-point correlation function (2PCF) and the power spectrum remain the conventional methods for analyzing LSS. The straightforward computation of these statistics and their sensitivity to the evolution and the growth of cosmic structure make them valuable tools for constraining cosmological parameters. So far, these methods have been applied with great success in a wide range of galaxy redshift surveys, such as the 2-degree Field Galaxy Redshift Survey (2dFGRS)~\cite{2dFGRS}, the 6-degree Field Galaxy Survey (6dFGS)~\cite{6dFGRS}, the WiggleZ survey~\cite{WiggleZ2011B,WiggleZ2011c}, and the Sloan Digital Sky Survey (SDSS)~\cite{SDSS_York,Eisenstein:2005su, Percival:2007yw,anderson2012clustering,sanchez2012clustering,sanchez2013clustering,anderson2014clustering,samushia2014clustering,ross2015clustering,beutler2016clustering,sanchez2016clustering,alam2017clustering,chuang2017clustering}, DESI~\cite{DESI}.


However, since the structure formation and evolution under gravitational interactions of matter gradually produce a non-Gaussian component, even though there is no non-Gaussianity in the cosmic primordial perturbation, such methods are sensitive only to the Gaussian part of the density field and cannot capture any non-Gaussian features or extract the relevant underlying cosmological information. To overcome this limitation and move beyond two-point statistics, numerous research efforts have been carried out, such as three-point statistics~\cite{Sabiu2016A&A,Slepian_2017}, four-point statistics~\cite{Sabiu_2019}, the study of cosmic voids~\cite{ryden1995measuring,lavaux2012precision}, and the application of deep learning~\cite{Ravanbakhsh17,Mathuriya18,pan2020cosmological,wang2023darkai}.

Besides, an alternative approach, the density-marked correlation function (MCF)~\cite{Beisbart:2000ja,Beisbart2002,Gottl2002,Sheth:2004vb,Sheth:2005aj,Skibba2006,White_2009,White2016,Satpathy:2019nvo,massara2020,Philcox2020}, has recently been proposed and proven to be an effective density statistic for the study of LSS in simulated data, with the advantage of computational simplicity. The MCF introduces a weight for each galaxy by using a "marker" that depends on its local density, thereby providing density-dependent clustering information. The traditional 2PCF approach is, of course, a special case of the MCF, where the weights are identical for all galaxies, regardless of differences in their physical properties and environments. What makes the MCF unique is its flexibility to emphasize statistics in higher- or lower-density regions, one can set the free-parameter weights to be proportional to positive or negative powers of the density. As demonstrated by~\cite{MCF_Yang} from the tests on mock data, assigning a density marker to galaxies and then weighting the labeled galaxies in combination with various weighting schemes can indeed extract more information beyond 2PCF.

In this study, we will, for the first time, assess the performance of the MCF in constraining cosmological parameters using real data. Specifically, we will apply the MCF to SDSS CMASS data based on a $\Lambda$CDM cosmology model, with a focus on constraining the matter density parameter, $\Omega_m$. The paper is structured as follows: in Sect.~\ref{sec:Data}, we introduce the observational data, SDSS BOSS DR12 CMASS galaxies, and a series of mocks used in our study. Sect.~\ref{sec:Methods} presents the methodology for calculating the MCF on both data and mocks, along with the constraint method employed. In Sect.~\ref{sec:Results}, we present our findings and compare them with constraints from previous research. Finally, we summarize our results and conclude in Sect.~\ref{sec:Conclusion}.

\section{Data}\label{sec:Data}

To derive cosmological constraints from MCF statistics, we need to compare actual observations with a series of simulations. Our analysis utilizes observations of SDSS BOSS DR12 CMASS galaxies~\cite{SDSS_Reid}. The analysis in this work relies on large-scale N-body simulations, including the BigMultiDark (BigMD)~\cite{rodriguez2016clustering_bigmd} and HR4 simulations~\cite{kim2015horizon, Jiang_2008_J08, Li16}, to account for potential systematic errors. Additionally, we use a series of fast simulations generated using the COmoving Lagrangian Acceleration (COLA) method~\cite{tassev2013_COLA}. In addition, we employ the Multidark Patchy simulations (referred to as Patchy)~\cite{kitaura2016clustering_Patchy} to estimate the relevant covariance.
 
In our analysis, we utilize the ROCKSTAR halo-finder~\cite{behroozi2012rockstar} in all simulations to detect gravitationally bound structures. ROCKSTAR operates through adaptive hierarchical refinement of friends-of-friends groups in six dimensions of phase space and one dimension of time, providing robust tracking of substructures. We consider both halos and subhalos in our analysis. Below, we provide a detailed description of the observational data and simulations used.

\subsection{Observational data}

\begin{figure}[H]
    \centering
    \includegraphics[width=0.5\textwidth]{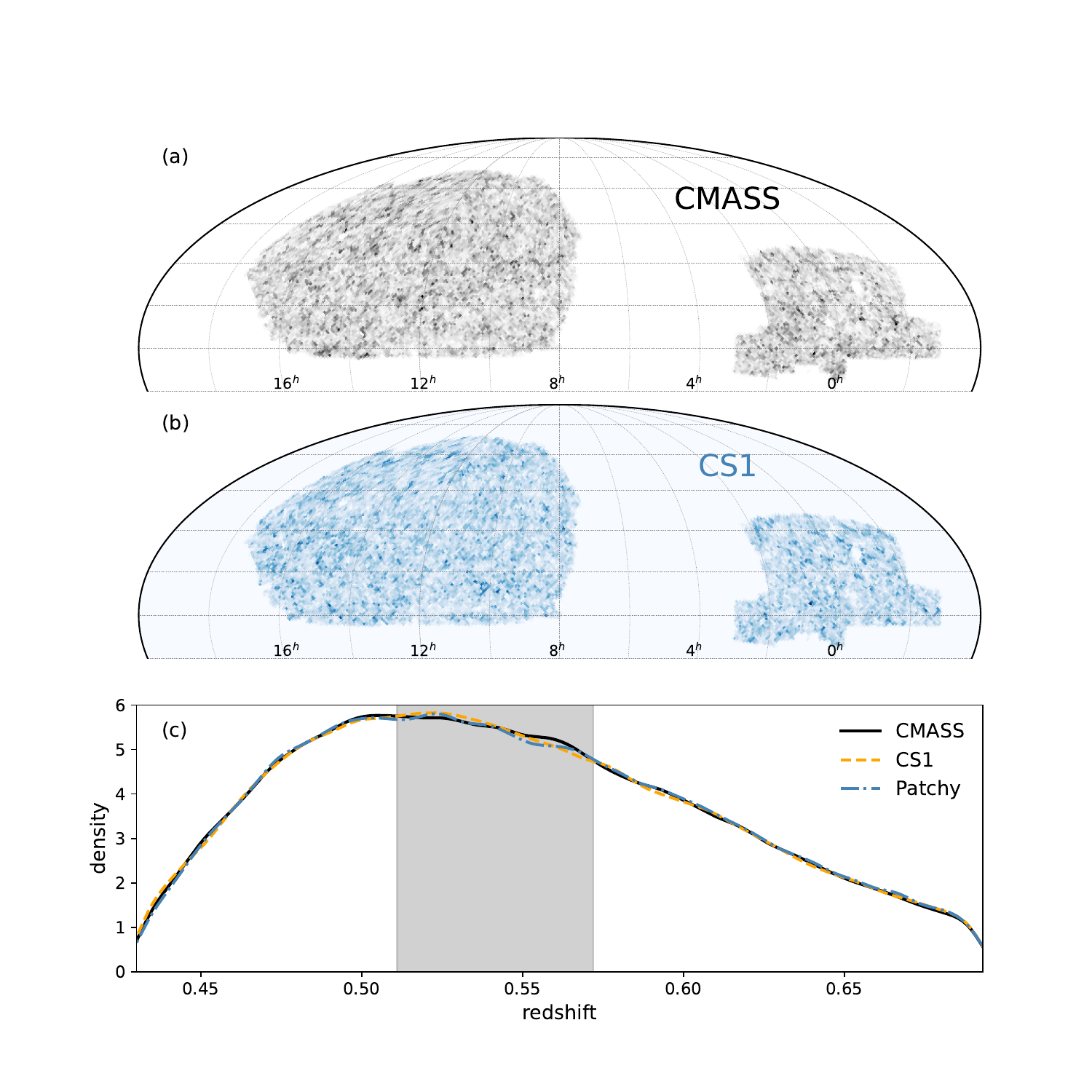}
    \caption{Spatial and redshift distributions of the observational galaxies and the simulation samples. Upper: the RA-DEC distribution of CMASS galaxies in redshift space within the range of $z \in [0.511, 0.572]$. Middle: the RA-DEC distribution of the generated CS1 mock sample, which includes both halos and subhalos and shares the same angular distribution as the CMASS galaxies. Bottom: the redshift distribution of CMASS (black solid line) and two mocks, CS1 (yellow dotted line) and Patchy (blue dashed-dotted line). The grey-shaded region represents the redshift range used in this study, i.e., $z \in [0.511, 0.572]$. The redshift distributions of the two mocks match well with that of CMASS.}
    \label{fig:radecz}
\end{figure}

The SDSS~\cite{SDSS_York} utilizes the 2.5-meter Sloan Telescope at the Apache Point Observatory in New Mexico for observations, primarily in five optical bands ($ugriz$)~\cite{gunn1998sloan,fukugita1996sloan}. The survey covers 7,606 and 3,172 square degrees in the northern and southern galactic hemispheres, respectively. As part of the SDSS-III project~\cite{eisenstein2011sdss}, BOSS~\cite{dawson2012baryon_BOSS,smee2013multi_BOSS} surveys a vast area of 9,376 square degrees. This covers 6,851 and 2,525 square degrees in the northern and southern hemispheres, respectively, as shown in Fig.~\ref{fig:radecz}. BOSS records spectra and redshifts for 1,372,737 galaxies, categorized into LOWZ ($z\lesssim0.4$) and CMASS ($0.4\lesssim z \lesssim0.6$) samples.

Our focus is on the CMASS sample, which targets massive, luminous galaxies with a mass threshold of approximately $M_*>10^{11} M_{\odot}$. Within the redshift range of $z\in[0.43,0.693]$, we have selected 771,567 galaxies, dividing them into three redshift bins: $[0.430,0.511]$, $[0.511, 0.572]$, and $[0.572,0.693]$.

From the CMASS and LOWZ observations, the average stellar number density is approximately $3\times10^{-4}$ $h^{3} {\rm Mpc}^{-3}$. To ensure comparability, we maintain this number as the halo number density in all our simulations. For CMASS galaxies with redshifts in the range of $z \in [0.511, 0.572]$, the spatial and redshift distributions are illustrated in Fig.~\ref{fig:radecz}.

To support the statistical analysis, the observation group released corresponding random samples. The angular and redshift selection criteria for these random samples are the same as for the actual observations, and the number of particles is 50 times higher than for the observations. When transforming a sample from angular and redshift distributions to comoving coordinates in this study, a flat $w$CDM cosmology model is assumed. In this model, the comoving distance is characterized by the present-day matter density parameter $\Omega_m$ and the equation of state of dark energy $w$. Specifically, our fiducial model is set with $\Omega_m = 0.26$ and $w=-1.0$, which is compatible with the WMAP5 cosmology.

When applying our fiducial cosmology model to convert redshifts to comoving distances for CMASS galaxies, the potential Alcock-Paczynski (AP) effect~\cite{LI14, LI15, Li16, LI18, LI19, Park:2019mvn, Zhang2019} may arise, resulting in redshift-dependent anisotropic clustering. To account for this effect consistently, we use the same redshift-distance relation for all simulation mocks in such conversion.

\subsection{COLA simulations}

In this study, in order to demonstrate the effectiveness of MCFs, we focus on considering the constraint on the cosmological parameter $\Omega_m$. To do so, in the COLA simulations, we systematically varied $\Omega_m$ in the range of $\Omega_m \in [0.2271, 0.3871]$, adjusting it in a small increment of $\Delta \Omega_m =0.005$. This results in a total of 34 changes. At the same time, the remaining cosmological parameters are adopted from the Planck measurements~\cite{collaboration2014planck},  with $\Omega_b=0.0482$, $\sigma_8=0.8228$, and $n_s=0.96$. To minimize the impact of sampling variance, we conducted 6 sets of COLA simulations, with each set comprising 34 simulations that varied in $\Omega_m$ (resulting in a total of $204$ simulations). However, we maintained the same initial random seed within each set. 
Therefore, the statistics presented in this paper for a given $\Omega_m$ are obtained by averaging 6 simulations with the same $\Omega_m$ in these sets.

Each COLA simulation includes approximately $1,200^3$ particles in a volume of $1,600^3$ $h^{-3}{\rm Mpc}^3$. To match the observations, we selected three snapshots, each with a redshift equal to the average value of the corresponding CMASS galaxy redshift bin.




\subsection{Correction data sets}

To correct various systematic errors in CMASS, our analysis relies on two sets of large N-body simulations, referred to as the Correction Sets (CS1 and CS2), which are detailed as follows.

\begin{itemize}
\item Correction Sets 1 (CS1):
the objective of CS1 is to correct the systematic effects induced by the mask, the radial variation of the number density, and the fiber collision effects observed in the data. For CS1, we utilized the mock survey of SDSS galaxies generated from the Horizon Run 4 (HR4) simulation~\cite{kim2015horizon,Jiang_2008_J08,Li16}. The mock survey has a similar angular distribution to CMASS galaxies, as demonstrated in Fig.~\ref{fig:radecz}. The CS1 dataset comprises 4 simulations with different initial random seeds, all constructed based on the $\Lambda$CDM cosmological model with the following parameters: $\Omega_m=0.26$, $\Omega_b=0.044$, $\sigma_8=0.85$, $n_s=0.96$, and $h=0.71$. The measurement statistics in CS1 are calculated by averaging over these 4 simulations to mitigate fluctuations arising from cosmic variance. Additionally, we consider the influence of galaxy mergers on cosmic evolution and calculate the timescales of these mergers with conventional theoretical predictions described in ~\cite{Jiang_2008_J08} to determine whether satellites completely disrupt. For this purpose, we use the J08 model~\cite{Jiang_2008_J08} as the default method for calculating merger timescales, enabling us to determine when satellite galaxies are completely destroyed. Note that the J08 model generates simulated galaxies that closely match observational data in terms of their properties and 2PCFs.
\end{itemize}

\begin{itemize}
\item Correction Sets 2 (CS2): to address the selection bias observed in CMASS galaxies, we conducted a separate series of CS2 simulations. These CS2 simulations were generated using the BigMultiDark (BigMD) simulation~\cite{rodriguez2016clustering_bigmd}, which employed $3840^3$ particles within a volume of $2500 {\rm h}^{-3} {\rm Mpc}^3$. The simulation assumes the $\Lambda$CDM cosmology: $\Omega_m=0.307115$, $\Omega_b=0.048206$, $\sigma_8=0.8288$, $n_s=0.96$, and $h=0.6777$. The substantial volume and particle number in CS2 adequately meet the requirements of our study. Furthermore, we incorporated the CMASS halo mass distribution~\cite{Berlind_2002_HOD} to accurately reproduce the selection bias observed in CMASS galaxies.
\end{itemize}



\subsection{Patchy simulation for covariance estimation}

To estimate measurement covariance matrices, we utilized the Patchy simulations~\cite{klypin2016multidark_patchy}, which contain approximately 2000 mock surveys. These simulations enable us to generate reliable estimates for the covariance. Patchy employs an approximate gravity solution and an analytical statistical bias model, under the same $\Lambda$CDM model parameters as in CS2 simulations. Notably, Patchy can successfully reproduce the stellar coefficient density, selection function, and observed geometry of BOSS DR12.
\section{Methodology}\label{sec:Methods}

In the following sections, we will provide a detailed description of the MCF statistical analysis and the strategies that we have employed.

MCF is a natural extension of the standard 2PCF, which is a statistical measure that involves assigning weights to objects before measuring the correlation between them~\cite{White2016,MCF_Yang,MCF_Xiao,Cai_2010_MCF,Chan_2022}. From a physical perspective, we assign weights to tracers based on their local density. This weighting approach helps us account for the environmental dependence in our analysis. This dependence arises because clustering and redshift space distortion effects vary significantly in dense and sparse regions of the universe. Following~\cite{MCF_Yang}, the weights are taken in the form of 
\begin{eqnarray}
    \label{eq:weightrho}
w(\boldsymbol{x})=\rho_{n_{\mathrm{NB}}}^\alpha \,.
\end{eqnarray}
Here the local density $\rho_{n_{\rm NB}}$ is estimated using its $n_{\rm NB}$ nearest neighbors,
\begin{eqnarray}
    \rho_{n_{\mathrm{NB}}}(\boldsymbol{r})=\sum_{i=1}^{n_{\mathrm{NB}}} W_k\left(\boldsymbol{r}-\boldsymbol{r}_i, h_W\right)\,,
\end{eqnarray}
where $\rho_{n_{\rm NB}}(\boldsymbol{r})$ is the environmental number density around a specific galaxy located at $\boldsymbol{r}$, and $W_k$ is the smoothing kernel, for which we choose the third-order B-spline
functions which has non-zero values within a sphere of radius $2 h_W$ $h^{-1} {\rm Mpc}$. The radius $h_w$ is determined by the 30-th nearest neighbor halos (i.e, $n_{\rm NB}=30$) ~\cite{Gingold1977,Lucy1977}. 

Compared with the traditional 2PCF, which is defined as $\xi(\boldsymbol{r})=\langle\delta(\boldsymbol{x}) \delta(\boldsymbol{x}+\boldsymbol{r})\rangle$, MCF takes the form of
\begin{eqnarray}
W(\boldsymbol{r})=\left\langle\delta(\boldsymbol{x}) \rho^\alpha_{n_{\mathrm{NB}}}(\boldsymbol{x})\, \delta(\boldsymbol{x}+\boldsymbol{r}) \rho^\alpha_{n_{\mathrm{NB}}}(\boldsymbol{x}+\boldsymbol{r})\right\rangle\,.
\end{eqnarray}
Note the distinction between $\delta$ and $\rho_{n_{\mathrm{NB}}}$. The latter represents the smoothed density field, whereas the former corresponds to the point-like density contrast, expressed as $\delta(\boldsymbol{x}) = \delta{\rho}/ \bar{\rho}$. Essentially, $\rho$ is a specific case of $\rho_{n_{\mathrm{NB}}}$ when $n_{\mathrm{NB}}=1$. Furthermore, when $\alpha=0$, the weights $\rho^\alpha=1$, causing the MCF to revert to the standard 2PCF.

Other than the weight assigned to each halo, the computational procedure for measuring the MCF is the same as that used for the standard 2PCFs. We employ the most commonly adopted Landy–Szalay estimator~\cite{Landy}: 
\begin{eqnarray}
    W(s,\mu) = \frac{WW-2WR+RR}{RR}\,,
\end{eqnarray}
where $WW$ represents the weighted number of galaxy-galaxy pairs, $WR$ corresponds to the galaxy-random pairs, and $RR$ denotes the random-random pairs. 
They are separated by a distance defined by $s \pm \Delta s$ and $\mu \pm \Delta \mu$, where $s$ represents the distance between pairs and $\mu$ is defined as $\cos(\theta)$, with $\theta$ being the angle between the line of sight (LOS) direction and the line connecting the pair. 
The random sample consists of 50 times the number of objects compared to the CMASS data and Patchy, and 40 times the number of objects compared to the CS1 simulations. Through our testing, we have determined that the results converge as long as the random sample contains at least 10 times more particles than both COLA and CS2 simulations. The weights of all particles are fixed to unity for the random samples.

\subsection{Two-dimensional MCFs: $\hat{W}(s,\mu)$}

It is common to use the normalized statistics, denoted as $\hat{W}$, to mitigate the effect of galaxy bias and to ensure a more accurate analysis. 
The normalized $\hat{W}(s,\mu)$ is defined as 
\begin{equation}\label{eq:norm}
    \hat W(s,\mu) = \frac{W(s,\mu)}{\int_{a}^{b} ds \int_0^{\mu_{\max}} W(s,\mu)d\mu}\,,
\end{equation}
where we choose $a = 20~h^{-1}{\rm Mpc}$, $b = 60~h^{-1}{\rm Mpc}$, and $\mu_{\max} = 0.97$. Through extensive testing, we find that the adopted parameters are sufficient to study clustering features effectively and perform well in distinguishing between different cosmologies.

\subsection{One-dimensional MCFs: $\hat W(s)$ and $\hat W_{\Delta s}(\mu)$  }

\begin{figure*}[htpb]
    \centering
    \includegraphics[width=1\textwidth]{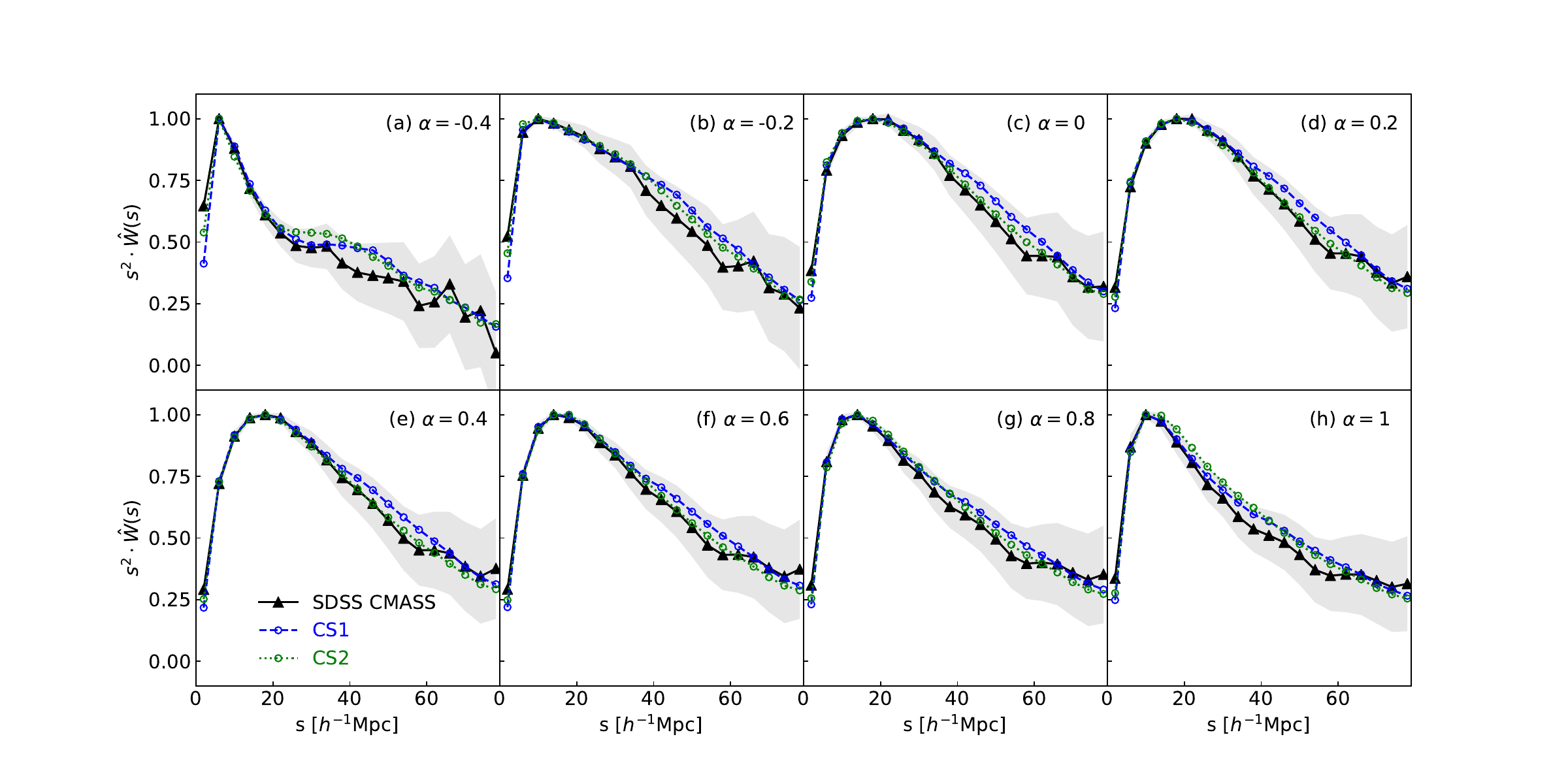}
    \caption{Comparison of one-dimensional MCFs, $s^2 \hat{W}(s)$, for three datasets: CMASS (red solid line), CS1 (blue dashed line), and CS2 (green dot-dashed line). To show the distribution of MCFs more clearly, we normalize the peak of each MCF to 1. The analysis is performed with parameter choice of the density weights $\alpha = [-0.4, -0.2, 0.0, 0.2, 0.4, 0.6, 0.8, 1.0]$, in the redshift range of $z \in [0.511, 0.572]$. 
    The grey-shaded region represents $2\sigma$ statistical errors estimated by using 2000 Patchy mocks. At different values of $\alpha$, the MCF distributions vary significantly. MCFs at $\alpha \in [0, 1]$ exhibit similar shapes but are notably different from the distribution at $\alpha = -0.4$. When increasing the absolute value of $|\alpha|$, it can be observed that the position of the peak shifts to the left, from 18 to 10 $h^{-1}~{\rm Mpc}$ and even at 6 $h^{-1}{\rm Mpc}$ for $\alpha=-0.4$. }
    \label{fig:xis}
\end{figure*}

\begin{figure*}[htpb]
    \centering
    \includegraphics[width=1\textwidth]{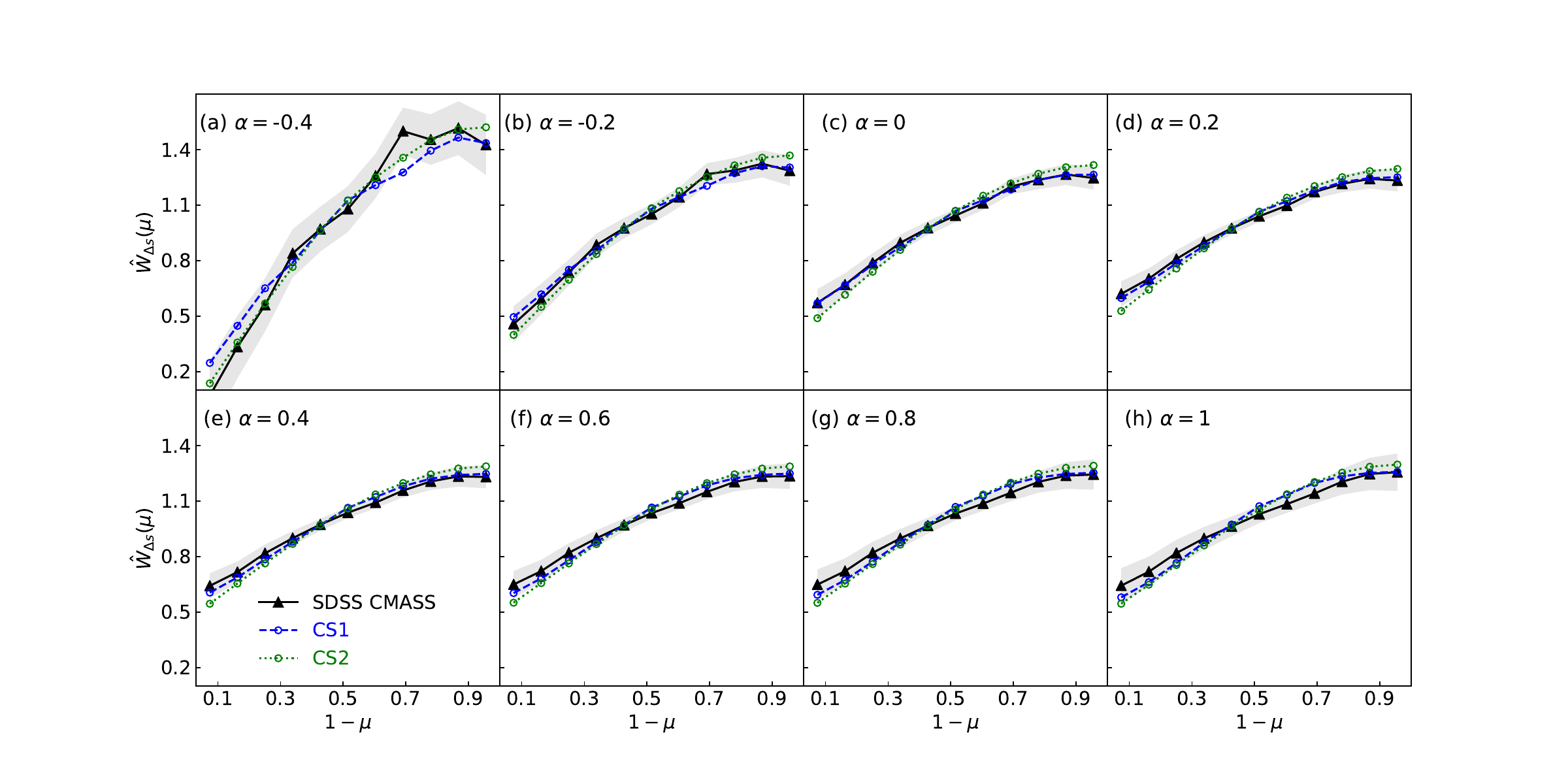}
    \caption{Same as in Fig.~\ref{fig:xis}, but for
    anisotropic clustering in the MCFs, $\hat{W}_{\Delta s}(\mu)$. The negative value of $\alpha=-0.4$ leads to distinctive behavior compared to $\alpha\geq 0$. For $\alpha\geq 0$, all the shapes are similar. Since we have chosen $\mu_{\max}=0.97$ in Eq.~\ref{eq:wsmu}, the FOG effect as $1-\mu$ approaches 0 is not apparent. In each panel, the amplitude positively depends on $1-\mu$, which is caused by the Kaiser effect.}
    \label{fig:ximu}
\end{figure*}

The RSDs in high-density and low-density regions exhibit significant differences. Consequently, we expect distinct anisotropic clustering features in the various MCFs. By performing integration of $W(s, \mu)$ over either the scale parameter ($s$) or the direction parameter ($\mu$), we can define two one-dimensional statistical quantities. The first one is the monopole of the MCF, which is a function of the clustering scale and is represented as
\begin{eqnarray}\label{eq:ws}
W_{\Delta \rm \mu}(s) = \int_{0}^{1} W(s, \mu)~d\mu\,.
\end{eqnarray}
The second quantity corresponds to the $\mu$-dependent function, and it is represented as
\begin{eqnarray}\label{eq:wmu}
W_{\Delta s}(\mu) = \int_{s_{\min }}^{s_{\max }} W(s, \mu)~ds\,,
\end{eqnarray}
where $s_{\rm min}=20~h^{-1}{\rm Mpc}$, $s_{\rm max}=40~h^{-1}{\rm Mpc}$, which have been employed to quantify both RSDs and AP distortions in the framework of the tomographic AP method~\cite{Li16}. In terms of Eqs.~\ref{eq:ws} and~\ref{eq:wmu}, the normalized quantities can be expressed as follows:
\begin{eqnarray}\label{eq:wsmu}
    \hat{W}_{\Delta \mu}(s) &=& 
    \frac{W_{\Delta \mu}(s)} {\int_a^b W_{\Delta \mu}(s)~ds}\ ,\nonumber \\ 
\hat{W}_{\Delta s}(\mu) &=& \frac{W_{\Delta s}(\mu)}{\int_0^{\mu_{\max}} W_{\Delta s}(\mu)~d \mu} \,,
\end{eqnarray}
where $a$, $b$ and $\mu_{\max}$ refer to the same values as those in Eq.~\ref{eq:norm}.


Fig.~\ref{fig:xis} displays the MCFs as functions of the clustering scale, specifically, the monopole $\hat{W}(s)$ here normalized by the peak of $s^2\hat{W}(s)$, with the direction dependence $\mu$ integrated out. We present the results using $\alpha = [-0.4, -0.2, 0.0, 0.2, 0.4, 0.6, 0.8, 1.0]$ in the redshift bin of $z \in [0.511,0.572]$, respectively. For all plots, we maintain a fixed value of $n_{\mathrm{NB}}=30$. A significant dependence on the weighting scheme becomes evident when comparing the MCFs obtained with different values of $\alpha$,  consistent with what was found in a previous study~\cite{MCF_Yang}. A larger $\alpha$ assigns greater weight to the dense, clustered regions, resulting in stronger correlations and thus presenting a distinct peak feature. This peak feature is not present in the standard 2PCF, offering additional information to constrain cosmological models.

Moreover, the clustering patterns in dense and underdense regions exhibit differences, leading to sensitivity in the shape of the MCFs with respect to $\alpha$. Additionally, when $\alpha=-0.4$, the shape of $\hat{W}(s)$ is distinct from those with positive $\alpha$. This is because the negative $\alpha$ will result in a large weight being assigned to particles in the underdense region. The distribution of $s^2\hat{W}(s)$ for $\alpha = -0.4$ (upper-left in Fig.~\ref{fig:xis}) at $s \in [10,60]$ $h^{-1}$ Mpc exhibits a relatively flat plateau shape. We also find that the peak gradually shifts towards zero when increasing the absolute value of $|\alpha|$. These $\alpha$-dependent features are consistent with results in the literature~\cite{MCF_Yang}.

Moreover, the gray shaded region corresponds to a $2\sigma$ statistical error, which was obtained by calculating the standard deviation of the MCF results from 2000 Patchy simulations. As seen, the resulting $\hat{W}(s)$ of CS1, CS2 and CMASS agree well within a $2\sigma$ level at the scales of $s\in[10,80]$ $h^{-1} \rm{Mpc}$.

Furthermore, in the range of $20-60$ $h^{-1}$ Mpc, the amplitudes for CS1 are generally higher than those of CS2 for most $\alpha$ values, except when $\alpha = -0.4$. However, the amplitudes of CS2 exceed those of CMASS for all $\alpha$ values, except also when $\alpha = -0.4$. This may be because the amplitude of the matter power spectrum in CS1 is slightly higher than that in CS2, with the former having $\sigma_8=0.85$ and the latter $\sigma_8=0.8288$. These results demonstrate the complex dependence of MCFs on $\alpha$, reflecting combined contributions from both higher- and lower-density regions. This complexity could potentially provide a means for discriminating between cosmological models.

Fig.~\ref{fig:ximu} shows the anisotropic clustering of MCFs, $\hat{W}_{\Delta s}(\mu)$, as defined in Eq.~\ref{eq:wsmu}. By varying $\alpha$ from -0.4 to 1, we observe a consistently positive slope across all cases. Notably, in the case of $\alpha=-0.4$, the slope becomes steeper. The positive dependence of the amplitude on $1-\mu$ is due to the Kaiser effect~\cite{Kaiser}. As observed, for each $\alpha$, the mocks of CS1 and CS2 present a similar shape for $\hat{W}_{\Delta s}(\mu)$ compared with CMASS, and all three are essentially compatible with each other within a $2\sigma$ level. However, a relatively large deviation is observed when $1-\mu<0.3$ among the simulations and observations, possibly due to the distinct impacts of the finger-of-god (FOG) effect.

\subsection{$\chi^2$ estimation}
We will perform the analysis using the statistical significance of variables, quantified using the chi-square ($\chi^2$) function. The estimation of the cosmological parameter $\Omega_m$ is performed by minimizing $\chi^2$ of the fit. The $\chi^2$ function is defined as
\begin{eqnarray}\label{eq:like}
\chi ^2 = (\boldsymbol{p}_{\rm model}-\boldsymbol{p}_{\rm data})^T \cdot \boldsymbol{C}^{-1} \cdot (\boldsymbol{p}_{\rm model}- \boldsymbol{p}_{\rm data})\,.
\end{eqnarray}
Here $\boldsymbol{p}_{\rm model}=\boldsymbol{p}_{\rm COLA}(\Omega_m)$ represents various statistical quantities of the mocks, including $\hat{W}(s)$, $\hat{W}_{\Delta s}(\mu)$, or $\hat W(s, \mu)$, and $\boldsymbol{C}$ denotes the covariance matrix of $\boldsymbol{p}$. Consequently, the $\chi^2$ values are used to constrain $\Omega_m$ by varying it within the COLA simulations.  Specifically, to correct for various systematic errors using CS1 and CS2 datasets, $\boldsymbol{p}_{\rm data}$ consists of three terms, which are 
\begin{eqnarray}\label{eq:correc1}
\boldsymbol{p}_{\rm data} = \boldsymbol{p}_{\rm CMASS} - \Delta \boldsymbol{p}_{\rm CS1} - \Delta \boldsymbol{p}_{\rm CS2}\,,
\end{eqnarray}
with 
\begin{eqnarray}\label{eq:correc2} 
\Delta \boldsymbol{p}_{\rm CS1} &=&\boldsymbol{p}_{\rm CS1}-\boldsymbol{p}_{\rm COLA}( \Omega_m=0.26)\,,\nonumber\\ 
\Delta \boldsymbol{p}_{\rm CS2}&=&\boldsymbol{p}_{\rm CS2}^{\rm CMASS}-\boldsymbol{p}_{\rm CS2}^{\rm COLA}\nonumber\,,
\end{eqnarray}
where the vector $\boldsymbol{p}$ represents various statistical quantities of CMASS data. It should be mentioned that the fiducial cosmology of $\boldsymbol{p}_{\rm COLA}( \Omega_m=0.26)$ is same with CS1. When combining MCFs with different density weights $\alpha$, we stack all measured statistics into the vector $\boldsymbol{p}$. The covariance matrices $\boldsymbol{C}$ for the integrated MCFs and the full shape of MCFs were estimated using a large number of Patchy mocks, and the details of this estimation will be provided in the following subsection.

Two additional terms, $\Delta \boldsymbol{p}_{\rm CS1}$ and $\Delta \boldsymbol{p}_{\rm CS2}$, are introduced for the following reasons~\cite{LI15}:
\begin{itemize}
    \item To correct for the systematic effects arising in CMASS, such as the impacts of the mask, the radial variation of the number density, and fiber collisions, we generate a COLA simulation with $\Omega_m$ identical to that in CS1. Thus, the term $\Delta \boldsymbol{p}_{\rm CS1}$ corresponds to the discrepancy between the COLA simulation and the actual observation, correcting for most of the observational effects in the CMASS data.

    \item
   The COLA simulations adopt a mass threshold to match the halo density with CMASS. However, the halo mass distributions between COLA and CMASS are not exactly the same. To address this selection bias, we generated two mock samples from CS2. One has the same mass function as that in CMASS, and another one the same as that in COLA. Therefore, the term $\Delta \boldsymbol{p}_{\rm CS2}$ provides the correction for the effect of the selection bias.
\end{itemize}

We have performed an analysis to compare the impact of including or excluding these corrections on the $\Omega_m$ value. The results, presented in Fig.~\ref{fig:barconstraints}, show a shift of $\Delta \Omega_m\approx 0.01$.

\subsection{Covariance estimation for $\hat W(s)$ and $\hat W_{\Delta s}(\mu)$}

\begin{figure*}[htpb]
	\centering
    \includegraphics[scale=0.55]{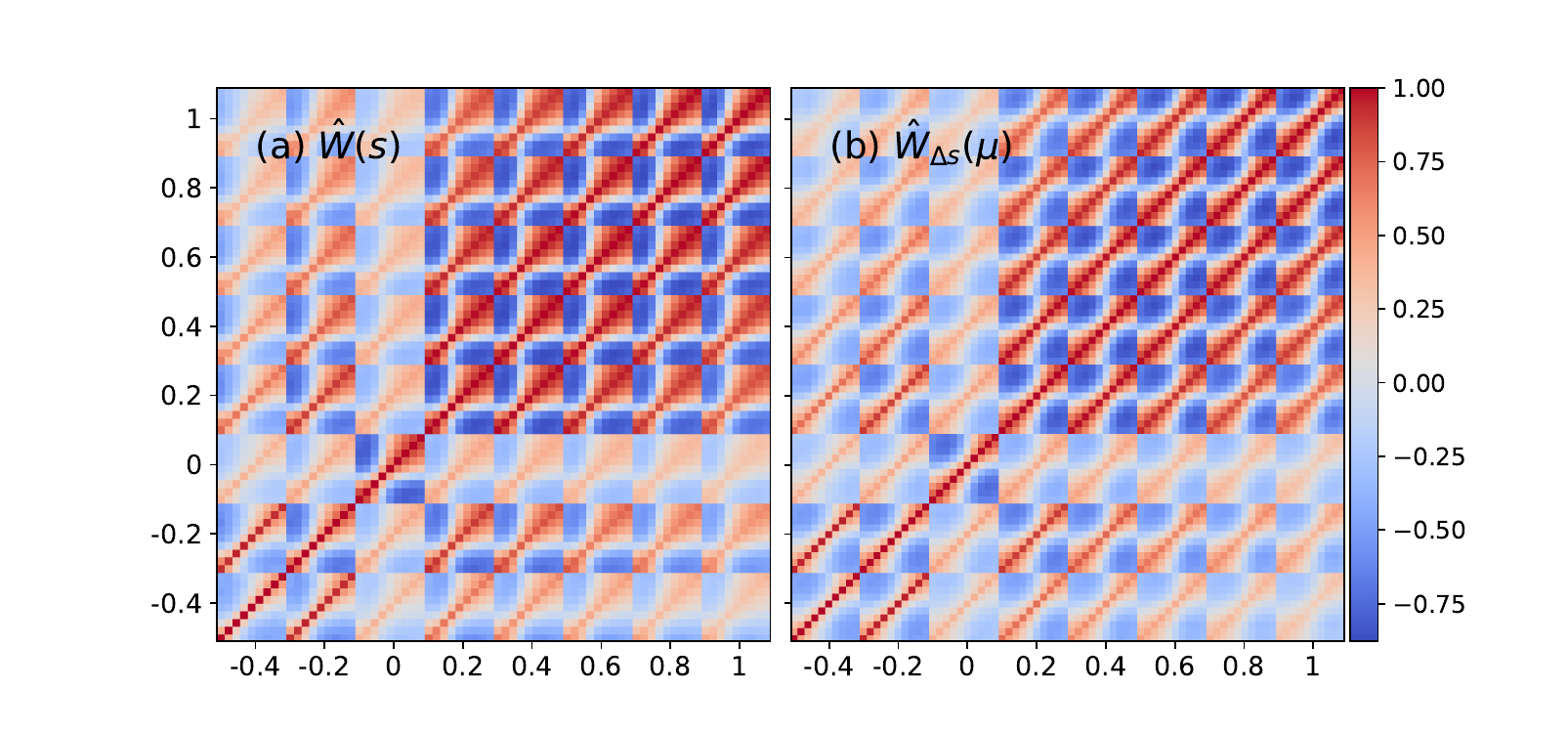}
    \caption{Correlation coefficient matrices for $\hat{W}(s)$ (left) and $\hat{W}_{\Delta s}(\mu)$ (right) with respect to $\alpha$, estimated using 2000 Patchy simulations. In total, we consider eight $\alpha$ values, namely $\alpha=[-0.4, -0.2, 0, 0.2, 0.4, 0.6, 0.8, 1]$ as indicated in the plots. Each block in the covariance matrices corresponds to the correlation coefficients for a given $\alpha$. These estimations are performed using the coarse bin scheme, where bin widths are fixed at $4~h^{-1} {\rm Mpc}$ and $0.083$, leading to 9 bins in $s$ and 10 bins in $\mu$, respectively. In both cases, the correlations among MCFs are strong when $\alpha \geq 0.2$, while the correlations with $\alpha \leq 0$ and those with other values of $\alpha$ weaken. Additionally, the correlations among MCFs with negative $\alpha$ are significantly smaller than those with $\alpha > 0.$}
    \label{fig:coef}
\end{figure*}

Our strategy is to perform a joint analysis using MCFs with different $\alpha$ values to place constraints on $\Omega_m$. Therefore, we need to compute the covariance matrix of different statistics across various $\alpha$ values. To estimate measurement covariance matrices accurately, we used 2000 Patchy simulations. We also note that the number of halos in each of the Patchy simulations, on average, is much higher than that in CMASS. To account for this, we introduced a rescale factor to enhance the Patchy-derived covariance matrix by a factor of $b_c=1.1$. This adjustment assumes that sampling variance is the major component in the covariance. Consequently, by utilizing the relationship that the covariance is inversely proportional to the number of halos, the value of $b_c$ is derived.



In Fig.~\ref{fig:coef},  we plot the corresponding correlation coefficient matrices ($r_{i j}\equiv C_{i j}/\sqrt{C_{i i} C_{j j}}$). This is done because the covariance matrices associated with distinct $\alpha$ values exhibit significantly varying magnitudes. The left and right panels illustrate the correlations of $\hat W(s)$ and $\hat W_{\Delta s}(\mu)$, respectively, with a focus on $\alpha=[-0.4, -0.2, 0, 0.2, 0.4, 0.6, 0.8, 1]$.

These covariance estimations are computed using the coarse bin scheme, wherein the bin widths are intentionally chosen to be relatively larger to ensure computational stability. The accurate estimation is performed for the finite number of Patchy mocks, with bin widths set at $4~h^{-1} {\rm Mpc}$ and $0.083$, resulting in 9 and 10 bins in $s$ and $\mu$, respectively.

For both correlation matrices, strong correlations are evident among MCFs when $\alpha> 0$, whereas correlations weaken when $\alpha\leq0$. Furthermore, it is important to note that correlations among MCFs with negative $\alpha$ are notably smaller compared to those with positive $\alpha$. Therefore, MCFs with negative $\alpha$ can offer supplementary information to complement the MCFs with the positive $\alpha$ values.

\subsection{Covariance estimation for $\hat W(s,\mu)$ and data compression}


In addition to performing cosmological analysis using the one-dimensional statistics $\hat W(s)$ or $\hat W_{\Delta s}(\mu)$, we also explore the two-dimensional statistic $\hat W(s, \mu)$. This full shape of $\hat W(s, \mu)$ can potentially provide more information, as some details could be lost during the integration of $s$ or $\mu$. 

\begin{figure}[H]
    \centering
    \includegraphics[width=0.5\textwidth]{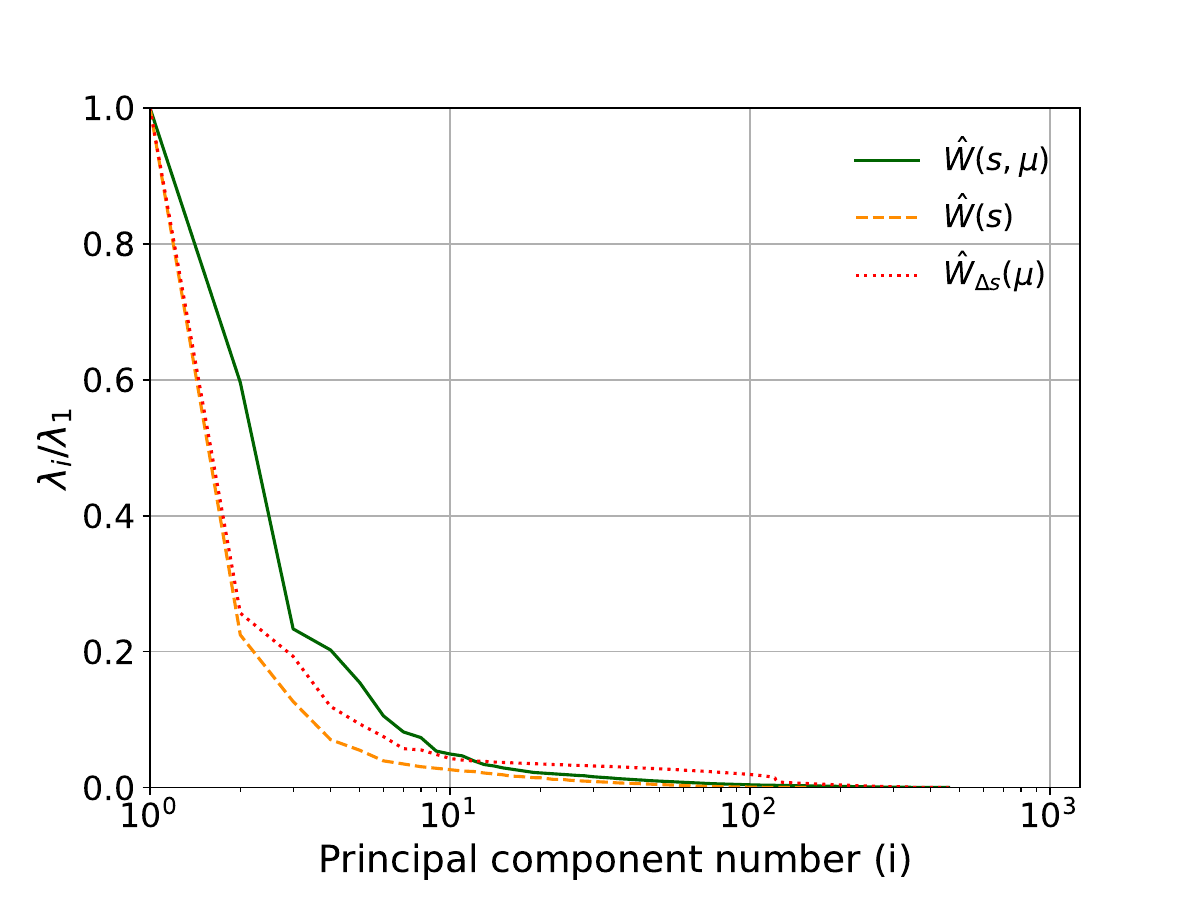}
    \caption{Ratio of the eigenvalue $\lambda_i$ with respect to the first one ($\lambda_1$) for MCFs $\hat{W}(s,\mu)$ (green), $\hat{W}(s)$ (yellow) and $\hat{W}_{\Delta s}(\mu)$ (red), which can be interpreted as the variance ratio by each eigenmode.  All the MCFs are obtained using a combination of weights $\alpha = [0,0.2,0.6]$.
    Based on the size of the covariance matrix, the total number of principal components is $1159$ for $\hat{W}(s,\mu)$, $39$ for $\hat{W}(s)$, and $116$ for $\hat{W}_{\Delta s}(\mu)$, respectively. It is evident that, in one-dimension case, the distribution of $\lambda_i$ experiences a sharp decrease until $\lambda_{10}$ and $\lambda_{20}$ for $\hat W(s,\mu)$, after which the eigenvalues of the remaining modes gradually decrease. The remaining modes may be dominated by noise, capturing only a small fraction of the information. Therefore, they can be safely ignored, which is why we choose the cutoff at $N_c=10$ for one-dimension MCF and $N_c=20$ for two-dimension MCF.}
    \label{fig:PCAeigenvalue}
\end{figure}

For a given $\alpha$, $\hat W(s, \mu)$ involves 117 $\mu$ bins and 40 $s$ bins, resulting in a total of $117 \times 40=460$ variables. Consequently, the number of variables in $\hat W(s, \mu)$ is notably large. In a joint analysis involving different $\alpha$ values, this situation can become even more challenging. 
One approach to address this challenge is to adopt the coarse bin scheme used in $\hat W(s)$ and $\hat W_{\Delta s}(\mu)$ for computational stability in covariance estimation, which can significantly reduce the number of variables. However, opting for coarse bins may result in a loss of information.

To numerically stabilize this estimation, we use Principal Component Analysis (PCA) to reduce the dimensionality of the statistics $\hat W(s,\mu)$  while preserving the majority of the cosmological information content. This PCA technique has already been used in cosmological analysis, as demonstrated in previous studies, such as~\cite{Petri_2016_PCA}.

PCA reduces the dimensionality of a dataset by transforming the data into a new basis. This orthogonal transformation converts a set of correlated variables into a set of uncorrelated variables by selecting several modes with the largest eigenvalues as the dimensionality reduction of the original data. PCA significantly reduces the number of variables while minimizing correlations among them, enhancing the feasibility and robustness of the analysis. It is assumed that only the statistical projections onto the first $N_c$ basis vectors contain relevant cosmological information, with the specific value of $N_c$ determined through simulations.

\begin{figure*}[htpb]
    \centering
    \includegraphics[scale=0.28]{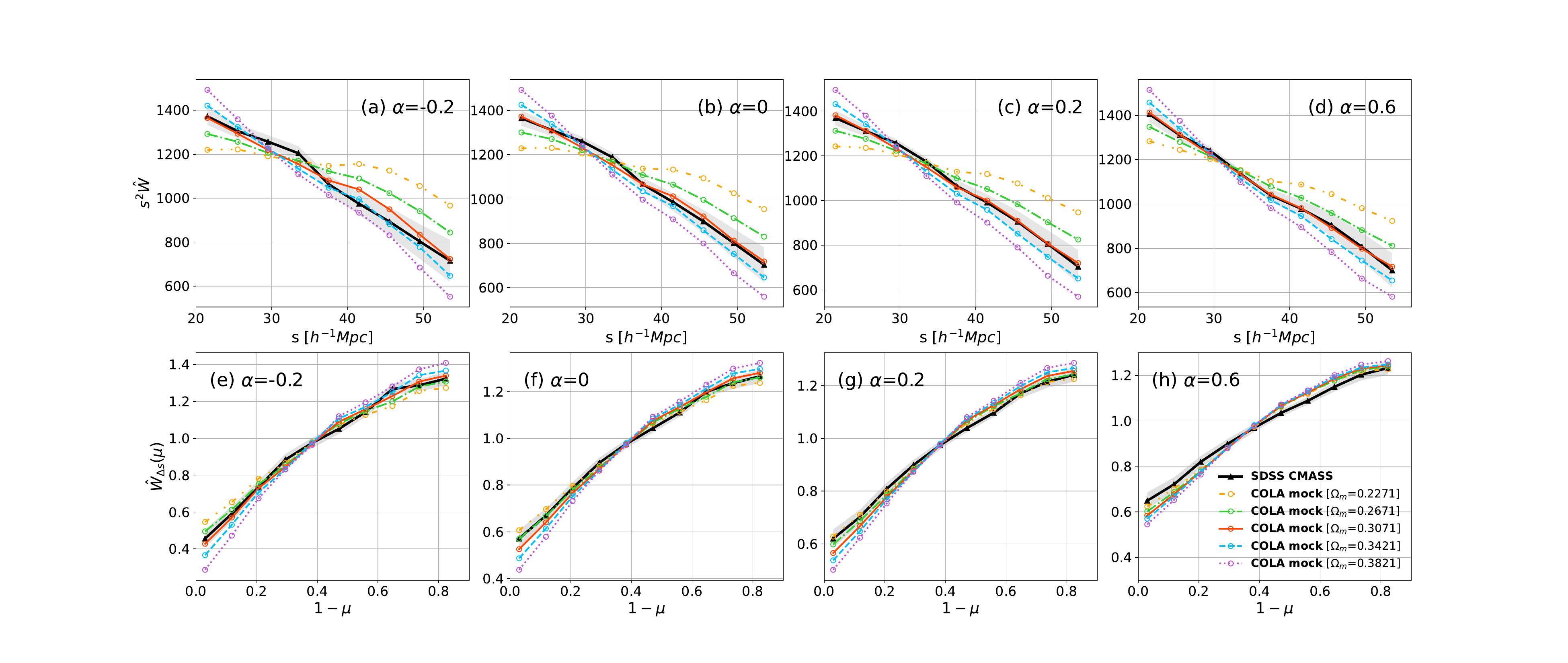}
    \caption{Comparison of the statistics of $\Delta\hat{W}(s)$ (upper panel) and $\Delta\hat{W}_{\Delta s}(\mu)$ (lower panel), as defined in Eq.~\ref{eq:wsmu}, at $\alpha = -0.2$ (left), $0$ (middle), and $0.6$ (right). By varying $\Omega_m$ in the range of $\Omega_m\in [0.227,0.382]$ for COLA simulations, the resulting MCFs are shown for comparison. In the lower panel, the integration is computed in the range of $s_{\rm min}=20$ $h^{-1}{\rm Mpc}$ and $s_{\rm max}=40$ $h^{-1}{\rm Mpc}$ for $\Delta\hat{W}_{\Delta s}(\mu)$. The systematic effects for CMASS have been corrected using the CS1 and CS2 simulation mocks, as described in Eq.~\ref{eq:correc1}. The corrected $\Delta\hat{W}(s)$ and $\Delta\hat{W}_{\Delta s}$ are shown in solid black. The grey-shaded region corresponds to the $2\sigma$ uncertainty estimated from 2000 Patchy mocks. Based on the test results, we only used MCFs within a specific range of values for $s$ and $\mu$, namely, $s\in[20, 60]$ $h^{-1}{\rm Mpc}$ and $1-\mu\in[0.03, 1]$, where the systematic effects are statistically not significant.}
    \label{fig:Mcf_diffcosmo}
\end{figure*} 

We perform PCA dimensionality reduction through the following procedure. We begin by estimating the $N_m \times N_m$ empirical covariance matrix of a vector $\boldsymbol{m}$, 
\begin{eqnarray}
\boldsymbol{C}^{\rm mock} &&= \left<\left(\boldsymbol{m}-\boldsymbol{\bar{m}}\right)\left(\boldsymbol{m}-\boldsymbol{\bar{m}}\right)^T \right> \\
&&=
\frac{1}{N_{\rm mock}-1} \sum_{i=1}^{N_{\rm mock}}  \left(\boldsymbol{m}_i - \boldsymbol{\bar{m}}\right) \left(\boldsymbol{m}_i-\boldsymbol{\bar{m}}\right)^T\,,
\end{eqnarray}
by averaging over all $N_{\rm mock}$ mock samples. Here, $\boldsymbol{m}_i$, with a length of $N_m$, denotes a vector containing all statistical quantities for the $i$-th mock sample. Note that the mean of $\boldsymbol{m}$ over all mocks is denoted by $\boldsymbol{\bar{m}}$.  We then diagonalize the covariance matrix $\boldsymbol{C}^{\rm mock}$ by performing a standard singular value decomposition (SVD),
\begin{eqnarray}
\boldsymbol{C}^{\rm mock} = \boldsymbol{U} \boldsymbol{\Lambda} \boldsymbol{U}^T\,,
\end{eqnarray}
where $\boldsymbol{U}$ is an orthogonal matrix with $\boldsymbol{U}\boldsymbol{U}^T = \boldsymbol{I}$, and its columns represent the eigenvectors. The matrix $\boldsymbol{\Lambda}$ is a diagonal matrix with elements $\Lambda_{jk} = \delta_{jk}\lambda_j$, where $\lambda_j$ are the eigenvalues sorted in decreasing order. By applying PCA dimensionality reduction to the data, we initially computed the covariance matrix $\boldsymbol{C}^{\rm mock}$ using the 2000 Patchy mocks. The distributions of eigenvalues for $\hat W(s, \mu)$, $\hat W(s)$ and $\hat W_{\Delta s}(\mu)$ of the mocks are illustrated in Fig.~\ref{fig:PCAeigenvalue}. We find that selecting $N_c=10$ is an appropriate choice for the one-dimension MCFs and $N_c=20$ for the two-dimension MCFs.

When applying the projection to the measured statistics, denoted as $\boldsymbol{m}'$ with a length of $N_m$ and keeping only the first $N_c$ dominant eigenmodes, we use the first $N_c$ columns of the matrix $\boldsymbol{U}$ for the projection. These selected columns are denoted as $\boldsymbol{U}_c$ and have dimensions $N_m \times N_c$ (where $N_c \ll N_m$).

This allows us to transform the data vector $\boldsymbol{m}'$ into the principal component (also known as the eigenmode) vector $\boldsymbol{a}$ using the projection:
\begin{eqnarray}\label{eq:pt}
\boldsymbol{a} = \boldsymbol{U}_c^T\boldsymbol{m}'\,,
\end{eqnarray}
where the length of $\boldsymbol{a}$ is reduced to $N_c$, significantly compressing the signal with negligible information loss.




The $\chi^2$ defined in Eq.~\ref{eq:like} can then be expressed as:
\begin{eqnarray}
\chi^2 = (\boldsymbol{p}_{\rm model}-\boldsymbol{p}_{\rm data})^T\cdot \boldsymbol{C}^{-1} \cdot (\boldsymbol{p}_{\rm model}-\boldsymbol{p}_{\rm data}) \approx  \sum_{i=1}^{N_c} \frac{a_i^2}{\lambda_i}\,,
\end{eqnarray} 
by using the transformation $ \boldsymbol{a} =\boldsymbol{U}^T_c (\boldsymbol{p}_{\rm model}-\boldsymbol{p}_{\rm data})$ by Eq.~\ref{eq:pt}. This transformation leads to a significant simplification and stabilization in the calculation. A further convergence test will be performed by varying $N_c$, and the results will be shown in Fig.~\ref{fig:pca_xismu}.

From the simulation tests and the amplitudes of the eigenvalues, we find this procedure can capture the cosmological information contained in $\hat W(s,\mu)$ by projecting it onto the $N_c=20$ principal components that exhibit the most significant variation with respect to $\Omega_m$. In the meantime, after performing PCA dimensionality reduction, the constraints in the $\Omega_m$ parameter space from the statistic can be as tight as those obtained with the full (pre-PCA) statistic vector. The stability and efficiency of the estimation based on pre-PCA statistics, however, are compromised.

\section{Results}\label{sec:Results}

In what follows, we will present the results and discuss the constraints on $\Omega_m$ using MCFs from the likelihood analysis. Additionally, we will provide a performance comparison between the standard 2PCF (i.e., the case of $\alpha=0$) and MCFs.

\subsection{Constraints from $\hat W(s)$ and $\hat W_{\Delta s}(\mu)$}

In Fig.~\ref{fig:Mcf_diffcosmo}, we illustrate the changes in $\hat W(s)$ and $\hat W_{\Delta s}(\mu)$ for different values of $\alpha = [-0.2, 0, 0.2, 0.6]$. In each panel, we provide the results for COLA mocks by varying $\Omega_m$ from 0.2271 to 0.3821, while we include a comparison with CMASS data. Concerning $\hat W(s)$, it is evident that significant changes occur when varying $\Omega_m$ for all $\alpha$ values. These changes can result in some results of COLA falling outside the $2\sigma$ range. likewise, $\hat W_{\Delta s}(\mu)$ is very   sensitive to $\Omega_m$, especially at $1-\mu\lesssim 0.4$ or $1-\mu\gtrsim0.8$. This suggests that the MCFs are very strongly dependent on the choice of the $\Omega_m$ value, with a high degree of discrimination. 

\begin{figure}[H]
    \centering
    \includegraphics[width=0.45\textwidth]{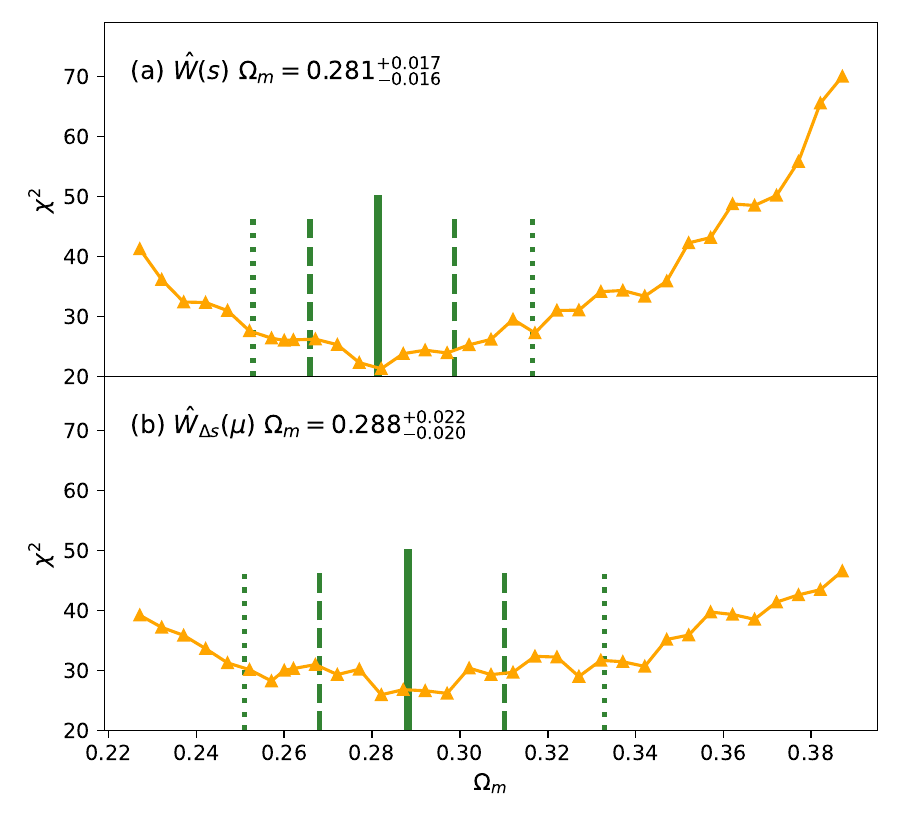}
    \caption{$\chi^2$ values for varying cosmological parameter $\Omega_m$, using the statistics $\hat{W}(s)$ (upper) and $\hat{W}_{\Delta s}(\mu)$ (lower). The $\chi^2$ results are derived from the joint analysis involving $\alpha = [0, 0.2, 0.6]$, using the coarse binning scheme. The median values are shown in solid green line, and the 68\% and 95\% confidence levels are depicted with dashed and dotted lines, respectively. The best-fit values are shown in red star scatter. The constraints at the $1\sigma$ level are $\Omega_m=0.281_{-0.016}^{+0.017}$ for $\hat{W}(s)$ and $\Omega_m=0.288_{-0.020}^{+0.022}$ for $\hat{W}_{\Delta s}(\mu)$, respectively.}
    \label{fig:chisquare}
\end{figure}

In Fig.~\ref{fig:chisquare}, we present the resulting $\chi^2$ values for $\hat{W}(s)$ (upper) and $\hat{W}_{\Delta s}(\mu)$ (lower) obtained through a joint analysis with $\alpha = [0, 0.2, 0.6]$ with using the coarse binning scheme. It is evident that the $\chi^2$ values increase rapidly when $\Omega_m$ deviates from approximately $0.28$. In comparison to $\hat{W}_{\Delta s}(\mu)$, the statistic $\hat{W}(s)$ offers slightly tighter constraints on $\Omega_m$, as there are more changes in $\chi^2$ when varying $\Omega_m$. Consequently, we derive  $\Omega_m=0.281_{-0.016}^{+0.017}$ from $\hat{W}(s)$ and $\Omega_m=0.288_{-0.020}^{+0.022}$ from $\hat{W}_{\Delta s}(\mu)$, at the $1\sigma$ level.

\subsection{Constraints from $\hat W(s,\mu)$}

\begin{figure}[H]
    \centering
    \includegraphics[scale=0.65]{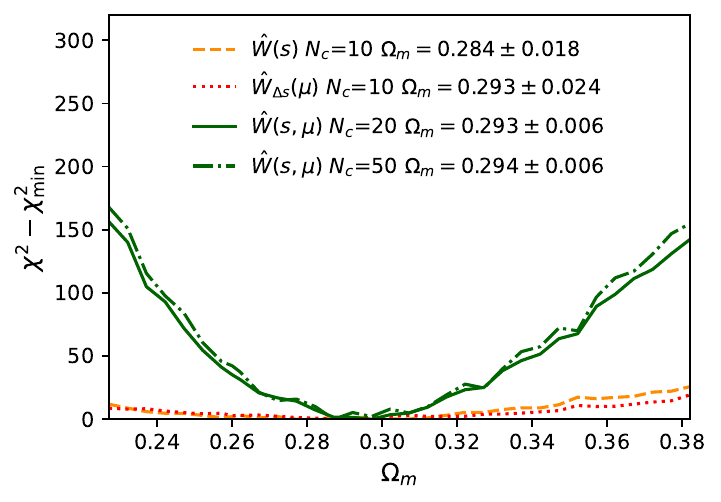}
    \caption{ $\chi^2$ values for $\hat{W}(s,\mu)$ (green), $\hat{W}(s)$ (yellow) and $\hat{W}_{\Delta s}(\mu)$ (red) as $\Omega_m$ varies, obtained through a joint analysis with $\alpha=[-0.2, 0, 0.2, 0.6]$. The technique of PCA dimensionality reduction has been applied in these analyses, with $N_c=10$ assumed as our default choice for one-dimension MCF and $N_c=20$ for two-dimension MCF. The two-dimensional statistic $\hat{W}(s,\mu)$ exhibits much stronger constraining power than the one-dimensional statistics, consistent with our expectation. For the convergence test, we increase the cut-off number for principal components from $N_c=20$ (green solid) to $N_c=50$ (green dot-dashed) for comparison. The changes in $\chi^2$ for these two $N_c$ values are compatible near the region of $\chi^2_{\min}$, hence providing similar constraints on $\Omega_m$. $\hat{W}(s,\mu)$ with $N_c=20$ yields $\Omega_m=0.293\pm{0.006}$ at the $1\sigma$ level, reducing the allowed region by about 66\% relative to those from the other integrated statistics.}
    \label{fig:pca_xismu}
\end{figure}

As expected, the two-dimensional statistic, $\hat W(s,\mu)$, contains more information than the integrated statistics. To further constrain $\Omega_m$, we performed an analysis by jointly combining various $\alpha$ values, specifically $\alpha=[-0.2, 0, 0.2, 0.6]$. Simultaneously, to stabilize the $\chi^2$ calculation, we employed the proposed PCA technique to reduce dimensionality. The orthogonal basis was constructed based on all of the Patch mocks. By projecting the measured $\hat W(s,\mu)$ vector from COLA onto the first $N_c=20$ basis vectors, the resulting $\chi^2$ values are presented in Fig.~\ref{fig:pca_xismu}. As observed, the allowed region is highly constrained since we have $117$ observables in $\hat{W}(s, \mu)$ for each $\alpha$ value, resulting in $\Omega_m=0.293\pm{0.006}$ at the $1\sigma$ level. The best-fit value is also $\Omega_m=0.293$, very close to the results obtained from either $\Omega_m =0.284$ for $\hat{W}(s)$  or  0.293 for $\hat{W}_{\Delta s}(\mu)$ alone. However, the error bar for $\hat W(s,\mu)$ is now reduced by $\sim 66\%$.

In addition, the application of PCA does indeed improve the stability of the calculation of $\chi^2$, as the changes in $\chi^2$ when varying $\Omega_m$ are smooth, with no strong fluctuations occurring. Furthermore, as a cross-check, we varied the number of independent eigenmodes from $N_c=4$ to $50$, and the results for $\hat W(s,\mu)$ are summarized as follows: $\Omega_m=0.279_{-0.008}^{+0.009}$ for $N_c=4$, $\Omega_m=0.287\pm{0.007}$ for $N_c=8$, and  $\Omega_m=0.284_{-0.006}^{+0.007}$ for $N_c=10$. Furthermore,   $\Omega_m=0.290\pm{0.006}$ for $N_c=14$, $0.293\pm{0.006}$ for $N_c=20$, $0.293\pm{0.006}$ for $N_c=30$, and $0.294\pm{0.006}$ for $N_c=50$. The estimates of $\Omega_m$ converge rapidly when $N_c \geq20$, demonstrating that $N_c = 20$ is an appropriate choice for adequately preserving information.


\subsection{Summary MCF results}\label{subsec:Comparison}

Let us first present a summary of the constraints on $\Omega_m$ in various cases. These results not only serve as a convergence test but also demonstrate the performance of each statistic. 

\begin{figure*}[tbph]
    \centering
    \includegraphics[scale=0.8]{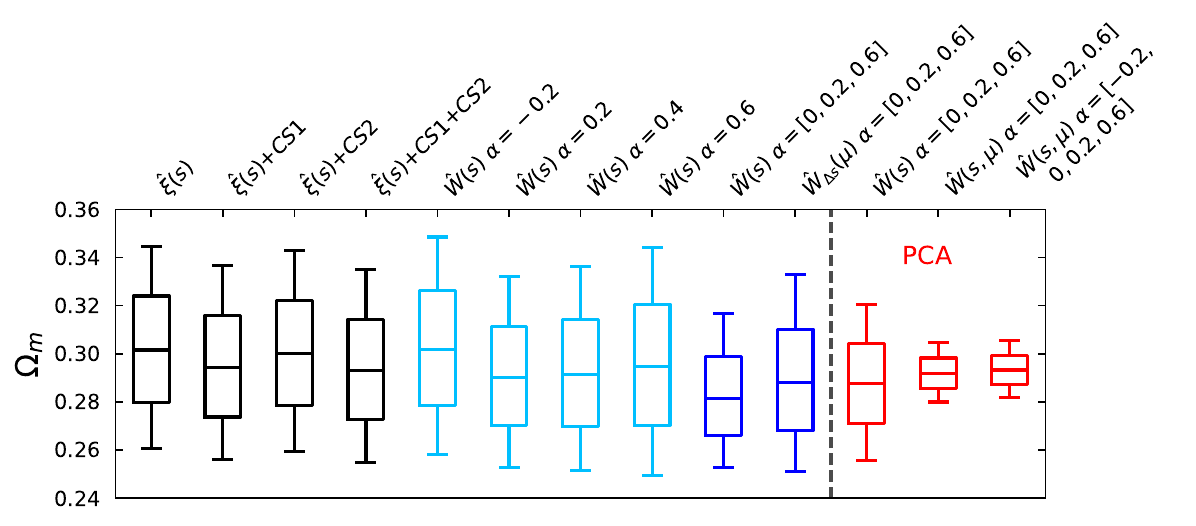}
    \caption{ Boxplot illustration of constraints on $\Omega_m$ for various cases, elaborately listed on the top $x$-axis. The $y$-axis represents the estimated values of $\Omega_m$ derived from the $\chi^2$ analysis. The boxes represent the $[16, 84]$ percentiles, while the vertical lines with horizontal edges represent the $[2.5, 97.5]$ percentiles, corresponding to the $1\sigma$ and $2\sigma$ intervals for the normal distribution. The central solid line within each box corresponds to the median value, while the green dashed line within each box corresponds to the best-fit value. All results have been obtained using 2000 Patch mocks to estimate errors. We emphasize that $\hat{W}(s, \mu)$, utilizing a combination of different weights, including $\alpha=[-0.2, 0, 0.2, 0.6]$, and employing PCA compression, provides the most stringent constraint. All the constraints on the left side of the black dotted line are derived using the coarse binning scheme.  }
    \label{fig:barconstraints}
\end{figure*}

In Fig.~\ref{fig:barconstraints}, we present both the median values and the best-fit values obtained by varying the weights $\alpha$ based on the CMASS data. We consider options including using $\hat W(s)$, $\hat W_{\Delta s}(\mu)$, or $\hat W(s,\mu)$ individually or in combination, with the coarse binning scheme or with PCA dimensionality reduction, while also assessing whether correction is implemented. Let us discuss the results one by one from left to right.

1) The first four results are obtained by using the statistic $\hat{\xi}(s)$ (i.e., $\hat W(s)$ for $\alpha=0$) while we consider the cases with and without CS1, and CS2 corrections respectively. As shown, these corrections shift the best-fit value of $\Omega_m$ at the $0.01$ level, while the statistical uncertainty remains unchanged because the corrections have no effect on the covariance estimates. Specifically, i) if we do not account for all corrections for the systematics, including the mask, angular incompleteness, radial variation, and fiber collision effect, then the estimated value is $\Omega_m=0.301\pm{0.022}$. ii) If we only consider the correction for the selection bias from CS2 mock but do not correct other bias from CS1, the derived constraint is $\Omega_m = 0.300\pm{0.022}$. iii) When we only take into account the correction from CS1, the constraint becomes slightly lower, yielding $\Omega_m = 0.294_{-0.021}^{+0.022}$. iv) If all systematics are debiased using the CS1 and CS2 simulation mocks, resulting in $\Omega_m=0.293_{-0.020}^{+0.021}$, this estimate is lower by $0.008$ compared to the case without any corrections.

2) Next, we present the results of one-dimensional MCFs, $\hat{W}(s), \hat{W}_{\Delta s}(\mu)$, after integrating over $s$ or $\mu$. We vary the density weight to demonstrate the dependence of the estimate on $\alpha$. It is noticeable that the error estimate for a specific $\alpha$ on its own is consistently much larger than the result obtained from a joint analysis involving different $\alpha$ values. This highlights the significance of performing a joint analysis across various $\alpha$ values (typically improving the constraint by $\sim 21\%$). Furthermore, the optimal values obtained are slightly sensitive to the choice of the statistics, and the variations are considerably smaller than the $1\sigma$ uncertainty. This implies that these estimations are robust.

3) The constraint on $\Omega_m$ can be further enhanced when using the complete vector of the 2D MCF $\hat{W}(s, \mu)$ because we incorporate all available statistical measurements, whereas the integrated MCFs result in information loss. It is evident that the errors are reduced by about $70\%$ and $63\%$ when comparing the results for $\hat{\xi}(s)$ and $\hat{W}(s)$ by combining $\alpha=[0, 0.2, 0.6]$, respectively. The strongest constraint in all of the above results arises from the use of $\hat{W}(s, \mu)$ in combination with all the considered weights, including $\alpha=[-0.2, 0, 0.2, 0.6]$. This estimate is $\Omega_m= 0.293\pm{0.006}$.  

\subsection{Comparison with other studies}\label{subsec:cother}
\begin{figure*}[htpb]
    \centering
    \includegraphics[scale=0.65]{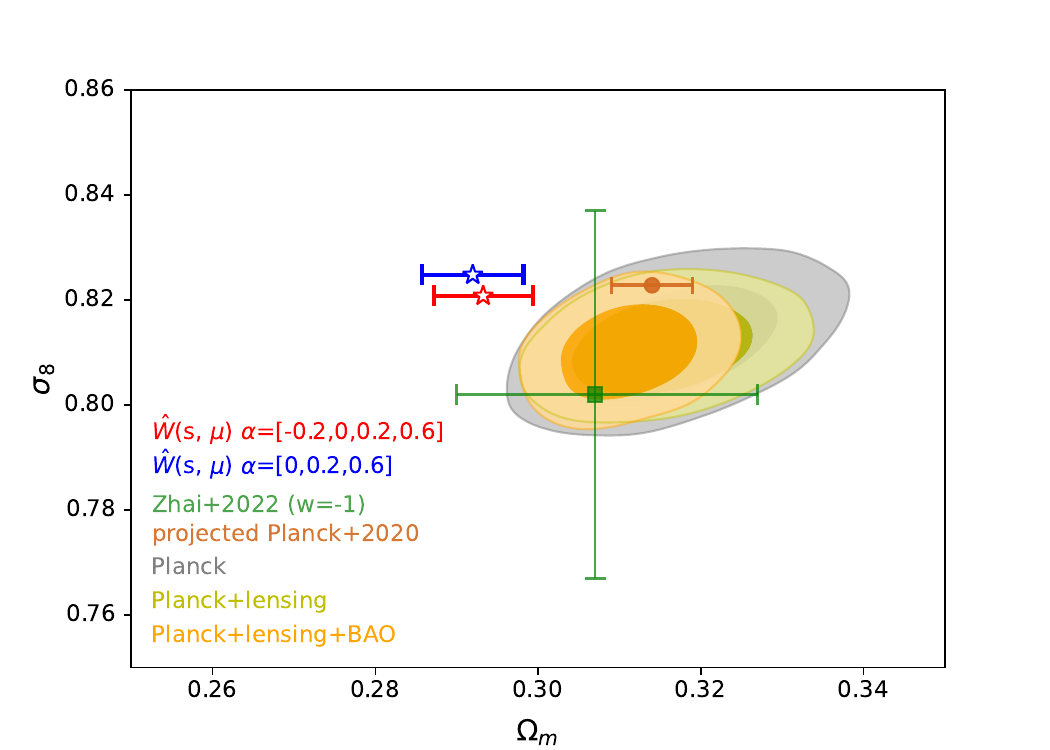}
    \caption{Measurements of $\Omega_m$ from our analysis using BOSS CMASS galaxies, along with a compilation of results from the literature. Our analysis results are based on the utilization of MCFs of $\hat{W}(s,\mu)$ with a combination of weights $\alpha=[-0.2, 0, 0.2, 0.6]$ (red) and $\alpha=[0, 0.2, 0.6]$ (blue), respectively. We have fixed $\sigma_8=0.8228$ in our analysis, resulting in $\Omega_m=0.293\pm{0.006}$ for $\alpha=[-0.2, 0, 0.2, 0.6]$ and $\Omega_m=0.292\pm{0.006}$ for $\alpha=[0,0.2,0.6]$. The shaded regions correspond to predictions at 68\% and 95\% confidence intervals from Planck together with lensing and BAO measurements~\cite{Planck2020}, assuming a flat $\Lambda$CDM cosmology. Keeping $\sigma_8$ fixed at the same value as used in our analysis for the Planck+Lensing+BAO measurements yields $\Omega_m=0.314\pm 0.005$. The recent estimate and its associated $1\sigma$ uncertainty ~\cite{zhai2022}, as derived from the small-scale clustering of BOSS galaxies, are shown in green, while assuming a fixed value of $w=-1$. 
    The estimates from MCFs are in good agreement with all other results. Notably, MCFs provide stronger constraints on $\Omega_m$, with the statistical error similar with that of the projected Planck measurement   (horizontal blue line). Note that, for the purpose of visualization, our measurements at the same $\sigma_8$ are slightly shifted.}
    \label{fig:resultcomparison}
\end{figure*}

In Fig.~\ref{fig:resultcomparison}, we present the results of constraints on $\Omega_m$ from the MCFs for $\hat{W}(s,\mu)$. These constraints are obtained by utilizing various weight combinations, specifically $\alpha=[0, 0.2, 0.6]$ (blue) and $\alpha=[-0.2, 0.2, 0, 0, 0.6]$ (red). Throughout our analysis, we maintain a fixed value of $\sigma_8=0.8228$. For the purpose of comparison, we include results from other studies in the literature, which include measurements\footnote{https://wiki.cosmos.esa.int/planck-legacy-archive/index.php/Cosmological\_Parameters/} from the Planck CMB data, large-scale structures including  the gravitational lensing (lensing), and the baryon acoustic oscillations (BAO)~\cite{Planck2020}. Additionally, we compare our findings with a recent result (green line)~\cite{zhai2022} derived from small-scale clustering measurements of BOSS galaxies, setting a $\Lambda$CDM model, utilizing a simulation-based emulator for 2PCF. By setting $\sigma_8$ to the same value as in our analysis, we project the posterior distribution for Planck+lensing+BAO, resulting in the constraint $\Omega_m=0.314\pm 0.005$ (blue line) at the $1\sigma$ level.

We can clearly observe that when using different combination of multiple $\alpha$ for MCFs, we obtain the constraints of $\Omega_m=0.293\pm{0.006}$ for $\alpha=[-0.2,0,0.2,0.6]$ and $\Omega_m=0.292\pm{0.006}$ for $\alpha=[0,0.2,0.6]$ , respectively. The statistical uncertainty of our results is comparable with the projected Planck result $\Omega_m=0.314\pm 0.005$. This suggests that MCFs can indeed provide a more comprehensive extraction of cosmological information underlying both high- and low-density regions. Additionally, our $\Omega_m$ values closely match that from small-scale clustering (yellow) within a $1\sigma$ deviation.
To quantify the tension between the MCFs and Planck measurements, under a Gaussian statistical assumption, the tension can be expressed as follows:
\begin{eqnarray}
T_{ij}=\frac{\left(\Omega_m^i-\Omega_m^j\right)}{\sqrt{\sigma_i^2+\sigma_j^2}}\,.
\end{eqnarray}
Here $i$ and $j$ represent the MCFs and Planck measurements, respectively. In addition, $\Omega_m$ denotes a median value of $\Omega_m$, and $\sigma$ represents the uncertainty associated with a measurement from either MCFs or Planck. Our MCFs-derived results agree well with recent simulation-based findings~\cite{zhai2022}, displaying a deviation of $0.7\sigma$ below their result for $\alpha=[-0.2,0,0.2,0.6]$ and $0.8\sigma$ below for $\alpha=[0,0.2,0.6]$. Moreover, we observe that our measurements are lower than the projected-Planck estimate by $2.6\sigma$ for $\alpha=[-0.2,0,0.2,0.6]$ and $2.8\sigma$ for $[0,0.2,0.6]$. It is evident that the $\Omega_m$ tension beyond the $2\sigma$ level exists between the measurements we obtained with the CMASS galaxies and that from the Planck observations under a flat $\Lambda$CDM cosmology.  While we acknowledge the potential existence of unknown systematic errors in the observations and galaxy models, our findings suggest the intriguing possibility of new physics beyond the standard cosmological model.


\section{Conclusion}\label{sec:Conclusion}

LSS surveys have increasingly offered stringent constraints on our cosmological models. A recent addition to this field is MCF, offering a computable density-correlation statistic. Simulations indicate that MCFs provide additional, independent constraints on cosmological models beyond the standard 2PCF. In this study, we have applied MCFs for the first time to SDSS CMASS data, focusing on utilizing three statistical quantities -- $\hat W(s)$, $\hat W_{\Delta s}(\mu)$, and $\hat{W}(s,\mu)$ -- for quantifying anisotropic clustering. Our aim is to explore statistical information regarding clustering and anisotropy properties in the universe, evaluating the performance of different weighting schemes in MCFs. Upon analyzing CMASS data, we have observed that combining different weights ($\alpha = [-0.2, 0, 0.2, 0.6]$) in MCFs provides the tight and independent constraint on the cosmological parameter $\Omega_m$. 

Additionally, we have introduced the PCA compression scheme, projecting extensive statistics into a few eigenmodes without a loss of information. In comparison to the coarse binning scheme, which adopts wider bins for variables, making the corresponding covariance matrices computationally acceptable, we find that choosing coarse bins could considerably weaken the constraint. The PCA compression scheme is particularly useful for the two-dimensional MCFs $\hat{W}(s,\mu)$ with a large number of variables. 

Importantly, the full shape of MCFs, $\hat{W}(s,\mu)$, obtained by combining multiple $\alpha$ values, could significantly reduce the statistical uncertainty of $\Omega_m$ by a factor of 2.7 to 3.4, compared to integrated MCFs or the standard 2PCF. Moreover, the most stringent constraints obtained from the $\alpha$-combined MCFs yield $\Omega_m=0.293\pm{0.006}$ for $\alpha=[-0.2,0,0.2,0.6]$. Our result is consistent with recent findings from simulation-based studies~\cite{zhai2022} within the 1$\sigma$ level. However, the estimated $\Omega_m$ deviates from the Planck results (with $\sigma_8$ fixed to our fiducial model) by approximately 0.021, indicating a tension in $\Omega_m$ about the $2.6\sigma$ level between CMASS and Planck measurements. 

Our study provides essential insights for maximizing the use of LSS information through MCFs, with the possibility of extending these methods to other surveys and datasets for constraining additional cosmological parameters. Indeed, it proves to be a valuable tool for upcoming emulator analyses on the Chinese Space Station Telescope (CSST). In the future, we also plan to implement enhancements based on the findings in~\cite{MCF_Xiao}, which indicate that incorporating the density gradient as a weight can offer additional information about LSS.

    





     

\section*{Acknowledgments}

We would like to thank Qinglin Ma and Yunlei Huang for their kind help. This work is supported by the Ministry of Science and Technology of China (2020SKA0110401, 2020SKA0110402, 2020SKA0110100), the National Key Research and Development Program of China (2018YFA0404504, 2018YFA0404601), the National Natural Science Foundation of China (11890691, 12205388, 12220101003, 12122301, 12233001), the China Manned Space Project with No. CMS-CSST-2021 (A02, A03, A04, B01), the Major Key Project of PCL, and the 111 project (B20019), and Shanghai Natural Science Research Grant (21ZR1430600). This work was performed using the Tianhe-2 supercomputer and the Kunlun cluster in the School of Physics and Astronomy at Sun Yat-Sen University. We also wish to acknowledge the Beijing Super Cloud Center (BSCC) and Beijing Beilong Super Cloud Computing Co., Ltd (http://www.blsc.cn/) for providing HPC resources that have significantly contributed to the research results presented in this paper. We thank the sponsorship from Yangyang Development Fund.

\section*{Data availability}

The SDSS BOSS DR12 CMASS galaxy catalog and the Patchy mocks, as well as their corresponding random samples, are available via the SDSS Science Archive Server\footnote{https://dr12.sdss.org/sas/dr12/boss/lss/}. The BigMD simulation used in this paper is available via the CosmoSim database\footnote{https://www.cosmosim.org/}.
The COLA simulation data and CS simulation sets used in this paper are available upon request. 


\InterestConflict{The authors declare that they have no conflict of interest.}

\bibliographystyle{scpma-zycai}
\bibliography{main}

\begin{thebibliography}{10}
\providecommand{\url}[1]{\texttt{#1}}
\providecommand{\urlprefix}{URL }
\providecommand{\doi}[1]{doi:~\href{http://doi.org/#1}{\nolinkurl{#1}}}
\providecommand{\arXiv}[1]{\href{https://arxiv.org/abs/#1}{\nolinkurl{https://arxiv.org/abs/#1}}}
\providecommand{\eprint}[1]{\href{http://arxiv.org/abs/#1}{\nolinkurl{#1}}}

\bibitem{zhai2022}
Z.~{Zhai}, J.~L. {Tinker}, A.~{Banerjee}, J.~{DeRose}, H.~{Guo}, Y.-Y. {Mao}, S.~{McLaughlin}, K.~{Storey-Fisher}, and R.~H. {Wechsler}, \href{http://dx.doi.org/10.48550/arXiv.2203.08999}{arXiv e-prints} arXiv:2203.08999 (2022), arXiv: \eprint{2203.08999}.

\bibitem{Riess_1998}
A.~G. Riess, A.~V. Filippenko, P.~Challis, A.~Clocchiatti, A.~Diercks, P.~M. Garnavich, R.~L. Gilliland, C.~J. Hogan, S.~Jha, R.~P. Kirshner, et~al., \href{http://dx.doi.org/10.1086/300499}{The Astronomical Journal} \textbf{116}, 1009 (1998), \urlprefix\url{https://iopscience.iop.org/article/10.1086/300499}.

\bibitem{Perlmutter_1999}
S.~Perlmutter, G.~Aldering, G.~Goldhaber, R.~A. Knop, P.~Nugent, P.~G. Castro, S.~Deustua, S.~Fabbro, A.~Goobar, D.~E. Groom, et~al., \href{http://dx.doi.org/10.1086/307221}{The Astrophysical Journal} \textbf{517}, 565 (1999), \urlprefix\url{https://iopscience.iop.org/article/10.1086/307221}.

\bibitem{Weinberg_1989}
S.~Weinberg, \href{http://dx.doi.org/10.1103/RevModPhys.61.1}{Rev. Mod. Phys.} \textbf{61}, 1 (1989), \urlprefix\url{https://link.aps.org/doi/10.1103/RevModPhys.61.1}.

\bibitem{Li_2011}
M.~Li, X.-D. Li, S.~Wang, and Y.~Wang, \href{http://dx.doi.org/10.1088/0253-6102/56/3/24}{Communications in Theoretical Physics} \textbf{56}, 525 (2011), \urlprefix\url{https://iopscience.iop.org/article/10.1088/0253-6102/56/3/24}.

\bibitem{YOO_2012}
J.~{Yoo} and Y.~{Watanabe}, \href{http://dx.doi.org/10.1142/S0218271812300029}{International Journal of Modern Physics D} \textbf{21}, 1230002 (2012), arXiv: \eprint{1212.4726}.

\bibitem{Weinberg_2013}
D.~H. {Weinberg}, M.~J. {Mortonson}, D.~J. {Eisenstein}, C.~{Hirata}, A.~G. {Riess}, and E.~{Rozo}, \href{http://dx.doi.org/10.1016/j.physrep.2013.05.001}{Physics Reports} \textbf{530}, 87 (2013), arXiv: \eprint{1201.2434}.

\bibitem{Bardeen_1986}
J.~M. {Bardeen}, J.~R. {Bond}, N.~{Kaiser}, and A.~S. {Szalay}, \href{http://dx.doi.org/10.1086/164143}{The Astrophysical Journal} \textbf{304}, 15 (1986).

\bibitem{deLapparent}
V.~{de Lapparent}, M.~J. {Geller}, and J.~P. {Huchra}, \href{http://dx.doi.org/10.1086/184625}{The Astrophysical Journall} \textbf{302}, L1 (1986).

\bibitem{Huchra}
J.~P. {Huchra}, L.~M. {Macri}, K.~L. {Masters}, T.~H. {Jarrett}, P.~{Berlind}, M.~{Calkins}, A.~C. {Crook}, R.~{Cutri}, P.~{Erdo{\v{g}}du}, E.~{Falco}, et~al., \href{http://dx.doi.org/10.1088/0067-0049/199/2/26}{The Astrophysical Journals} \textbf{199}, 26 (2012), arXiv: \eprint{1108.0669}.

\bibitem{Tegmark}
M.~Tegmark, M.~Blanton, M.~Strauss, F.~Hoyle, D.~Schlegel, R.~Scoccimarro, M.~Vogeley, D.~Weinberg, I.~Zehavi, A.~Berlind, et~al., \href{http://dx.doi.org/10.1086/382125}{Astrophysical Journal} \textbf{606}, 702 (2004).

\bibitem{Guzzo}
L.~{Guzzo}, M.~{Scodeggio}, B.~{Garilli}, B.~R. {Granett}, A.~{Fritz}, U.~{Abbas}, C.~{Adami}, S.~{Arnouts}, J.~{Bel}, M.~{Bolzonella}, et~al., \href{http://dx.doi.org/10.1051/0004-6361/201321489}{Astronomy and Astrophysics} \textbf{566}, A108 (2014), arXiv: \eprint{1303.2623}.

\bibitem{wu2023prospects}
P.-J. Wu, Y.~Li, J.-F. Zhang, and X.~Zhang, Science China Physics, Mechanics \& Astronomy \textbf{66}, 270413 (2023).

\bibitem{xu2020cosmological}
Y.~Xu and X.~Zhang, arXiv preprint arXiv:2002.00572  (2020).

\bibitem{zhang2023cosmology}
J.-G. Zhang, Z.-W. Zhao, Y.~Li, J.-F. Zhang, D.~Li, and X.~Zhang, Science China Physics, Mechanics \& Astronomy \textbf{66}, 120412 (2023).

\bibitem{armano2016sub}
M.~Armano, H.~Audley, G.~Auger, J.~T. Baird, M.~Bassan, P.~Binetruy, M.~Born, D.~Bortoluzzi, N.~Brandt, M.~Caleno, et~al., Physical review letters \textbf{116}, 231101 (2016).

\bibitem{hu2017taiji}
W.-R. Hu and Y.-L. Wu, The taiji program in space for gravitational wave physics and the nature of gravity (2017).

\bibitem{luo2020first}
J.~Luo, Y.-Z. Bai, L.~Cai, B.~Cao, W.-M. Chen, Y.~Chen, D.-C. Cheng, Y.-W. Ding, H.-Z. Duan, X.~Gou, et~al., Classical and Quantum Gravity \textbf{37}, 185013 (2020).

\bibitem{wang2022forecast}
L.-F. Wang, S.-J. Jin, J.-F. Zhang, and X.~Zhang, Science China Physics, Mechanics \& Astronomy \textbf{65}, 210411 (2022).

\bibitem{EUCLID}
R.~{Laureijs}, J.~{Amiaux}, S.~{Arduini}, J.~L. {Augu{\`e}res}, J.~{Brinchmann}, R.~{Cole}, M.~{Cropper}, C.~{Dabin}, L.~{Duvet}, and A.~{Ealet}, \href{https://ui.adsabs.harvard.edu/abs/2011arXiv1110.3193L}{arXiv e-prints} arXiv:1110.3193 (2011), arXiv: \eprint{1110.3193}.

\bibitem{LSST}
{LSST Science Collaboration}, P.~A. {Abell}, J.~{Allison}, S.~F. {Anderson}, J.~R. {Andrew}, J.~R.~P. {Angel}, L.~{Armus}, D.~{Arnett}, S.~J. {Asztalos}, and T.~S. {Axelrod}, \href{https://ui.adsabs.harvard.edu/abs/2009arXiv0912.0201L}{arXiv e-prints} arXiv:0912.0201 (2009), arXiv: \eprint{0912.0201}.

\bibitem{WFIRST}
D.~{Spergel}, N.~{Gehrels}, C.~{Baltay}, D.~{Bennett}, J.~{Breckinridge}, M.~{Donahue}, A.~{Dressler}, B.~S. {Gaudi}, T.~{Greene}, and O.~{Guyon}, \href{https://ui.adsabs.harvard.edu/abs/2015arXiv150303757S}{arXiv e-prints} arXiv:1503.03757 (2015), arXiv: \eprint{1503.03757}.

\bibitem{Gong_2019}
Y.~Gong, X.~Liu, Y.~Cao, X.~Chen, Z.~Fan, R.~Li, X.-D. Li, Z.~Li, X.~Zhang, and H.~Zhan, \href{http://dx.doi.org/10.3847/1538-4357/ab391e}{The Astrophysical Journal} \textbf{883}, 203 (2019), \urlprefix\url{https://iopscience.iop.org/article/10.3847/1538-4357/ab391e}.

\bibitem{li2023forecast}
S.-Y. Li, Y.-L. Li, T.~Zhang, J.~Vink{\'o}, E.~Reg{\H{o}}s, X.~Wang, G.~Xi, and H.~Zhan, Science China Physics, Mechanics \& Astronomy \textbf{66}, 229511 (2023).

\bibitem{Kaiser}
N.~{Kaiser}, \href{http://dx.doi.org/10.1093/mnras/227.1.1}{Monthly Notices of the Royal Astronomical Society} \textbf{227}, 1 (1987).

\bibitem{Ballinger}
W.~E. {Ballinger}, J.~A. {Peacock}, and A.~F. {Heavens}, \href{http://dx.doi.org/10.1093/mnras/282.3.877}{Monthly Notices of the Royal Astronomical Society} \textbf{282}, 877 (1996), arXiv: \eprint{astro-ph/9605017}.

\bibitem{Eisenstein_1998}
D.~J. {Eisenstein} and W.~{Hu}, \href{http://dx.doi.org/10.1086/305424}{The Astrophysical Journal} \textbf{496}, 605 (1998), arXiv: \eprint{astro-ph/9709112}.

\bibitem{Blake_2003}
C.~{Blake} and K.~{Glazebrook}, \href{http://dx.doi.org/10.1086/376983}{The Astrophysical Journal} \textbf{594}, 665 (2003), arXiv: \eprint{astro-ph/0301632}.

\bibitem{Seo_2003}
H.-J. {Seo} and D.~J. {Eisenstein}, \href{http://dx.doi.org/10.1086/379122}{The Astrophysical Journal} \textbf{598}, 720 (2003), arXiv: \eprint{astro-ph/0307460}.

\bibitem{2dFGRS}
M.~{Colless}, B.~A. {Peterson}, C.~{Jackson}, J.~A. {Peacock}, S.~{Cole}, P.~{Norberg}, I.~K. {Baldry}, C.~M. {Baugh}, J.~{Bland-Hawthorn}, T.~{Bridges}, et~al., \href{http://dx.doi.org/10.48550/arXiv.astro-ph/0306581}{arXiv e-prints} astro-ph/0306581 (2003), arXiv: \eprint{astro-ph/0306581}.

\bibitem{6dFGRS}
F.~Beutler, C.~Blake, M.~Colless, D.~H. Jones, L.~Staveley-Smith, G.~B. Poole, L.~Campbell, Q.~Parker, W.~Saunders, and F.~Watson, \href{http://dx.doi.org/10.1111/j.1365-2966.2012.21136.x}{Monthly Notices of the Royal Astronomical Society} \textbf{423}, 3430 (2012), arXiv: \eprint{https://academic.oup.com/mnras/article-pdf/423/4/3430/4903419/mnras0423-3430.pdf}, \urlprefix\url{https://doi.org/10.1111/j.1365-2966.2012.21136.x}.

\bibitem{WiggleZ2011B}
C.~{Blake}, K.~{Glazebrook}, T.~M. {Davis}, S.~{Brough}, M.~{Colless}, C.~{Contreras}, W.~{Couch}, S.~{Croom}, M.~J. {Drinkwater}, K.~{Forster}, et~al., \href{http://dx.doi.org/10.1111/j.1365-2966.2011.19606.x}{Monthly Notices of the Royal Astronomical Society} \textbf{418}, 1725 (2011), arXiv: \eprint{1108.2637}.

\bibitem{WiggleZ2011c}
C.~{Blake}, S.~{Brough}, M.~{Colless}, C.~{Contreras}, W.~{Couch}, S.~{Croom}, T.~{Davis}, M.~J. {Drinkwater}, K.~{Forster}, D.~{Gilbank}, et~al., \href{http://dx.doi.org/10.1111/j.1365-2966.2011.18903.x}{Monthly Notices of the Royal Astronomical Society} \textbf{415}, 2876 (2011), arXiv: \eprint{1104.2948}.

\bibitem{SDSS_York}
D.~G. {York}, J.~{Adelman}, J.~{Anderson}, John~E., S.~F. {Anderson}, J.~{Annis}, N.~A. {Bahcall}, J.~A. {Bakken}, R.~{Barkhouser}, S.~{Bastian}, E.~{Berman}, et~al., \href{http://dx.doi.org/10.1086/301513}{The Astronomical Journal} \textbf{120}, 1579 (2000), arXiv: \eprint{astro-ph/0006396}.

\bibitem{Eisenstein:2005su}
D.~J. {Eisenstein}, I.~{Zehavi}, D.~W. {Hogg}, R.~{Scoccimarro}, M.~R. {Blanton}, R.~C. {Nichol}, R.~{Scranton}, H.-J. {Seo}, M.~{Tegmark}, Z.~{Zheng}, et~al., \href{http://dx.doi.org/10.1086/466512}{The Astrophysical Journal} \textbf{633}, 560 (2005), arXiv: \eprint{astro-ph/0501171}.

\bibitem{Percival:2007yw}
W.~J. {Percival}, S.~{Cole}, D.~J. {Eisenstein}, R.~C. {Nichol}, J.~A. {Peacock}, A.~C. {Pope}, and A.~S. {Szalay}, \href{http://dx.doi.org/10.1111/j.1365-2966.2007.12268.x}{Monthly Notices of the Royal Astronomical Society} \textbf{381}, 1053 (2007), arXiv: \eprint{0705.3323}.

\bibitem{anderson2012clustering}
L.~{Anderson}, E.~{Aubourg}, S.~{Bailey}, D.~{Bizyaev}, M.~{Blanton}, A.~S. {Bolton}, J.~{Brinkmann}, J.~R. {Brownstein}, A.~{Burden}, A.~J. {Cuesta}, et~al., \href{http://dx.doi.org/10.1111/j.1365-2966.2012.22066.x}{Monthly Notices of the Royal Astronomical Society} \textbf{427}, 3435 (2012), arXiv: \eprint{1203.6594}.

\bibitem{sanchez2012clustering}
A.~G. {S{\'a}nchez}, C.~G. {Sc{\'o}ccola}, A.~J. {Ross}, W.~{Percival}, M.~{Manera}, F.~{Montesano}, X.~{Mazzalay}, A.~J. {Cuesta}, D.~J. {Eisenstein}, E.~{Kazin}, et~al., \href{http://dx.doi.org/10.1111/j.1365-2966.2012.21502.x}{Monthly Notices of the Royal Astronomical Society} \textbf{425}, 415 (2012), arXiv: \eprint{1203.6616}.

\bibitem{sanchez2013clustering}
A.~G. {S{\'a}nchez}, E.~A. {Kazin}, F.~{Beutler}, C.-H. {Chuang}, A.~J. {Cuesta}, D.~J. {Eisenstein}, M.~{Manera}, F.~{Montesano}, R.~C. {Nichol}, N.~{Padmanabhan}, et~al., \href{http://dx.doi.org/10.1093/mnras/stt799}{Monthly Notices of the Royal Astronomical Society} \textbf{433}, 1202 (2013), arXiv: \eprint{1303.4396}.

\bibitem{anderson2014clustering}
L.~{Anderson}, {\'E}.~{Aubourg}, S.~{Bailey}, F.~{Beutler}, V.~{Bhardwaj}, M.~{Blanton}, A.~S. {Bolton}, J.~{Brinkmann}, J.~R. {Brownstein}, A.~{Burden}, et~al., \href{http://dx.doi.org/10.1093/mnras/stu523}{Monthly Notices of the Royal Astronomical Society} \textbf{441}, 24 (2014), arXiv: \eprint{1312.4877}.

\bibitem{samushia2014clustering}
L.~{Samushia}, B.~A. {Reid}, M.~{White}, W.~J. {Percival}, A.~J. {Cuesta}, G.-B. {Zhao}, A.~J. {Ross}, M.~{Manera}, {\'E}.~{Aubourg}, F.~{Beutler}, et~al., \href{http://dx.doi.org/10.1093/mnras/stu197}{Monthly Notices of the Royal Astronomical Society} \textbf{439}, 3504 (2014), arXiv: \eprint{1312.4899}.

\bibitem{ross2015clustering}
A.~J. {Ross}, L.~{Samushia}, C.~{Howlett}, W.~J. {Percival}, A.~{Burden}, and M.~{Manera}, \href{http://dx.doi.org/10.1093/mnras/stv154}{Monthly Notices of the Royal Astronomical Society} \textbf{449}, 835 (2015), arXiv: \eprint{1409.3242}.

\bibitem{beutler2016clustering}
F.~{Beutler}, H.-J. {Seo}, A.~J. {Ross}, P.~{McDonald}, S.~{Saito}, A.~S. {Bolton}, J.~R. {Brownstein}, C.-H. {Chuang}, A.~J. {Cuesta}, D.~J. {Eisenstein}, et~al., \href{http://dx.doi.org/10.1093/mnras/stw2373}{Monthly Notices of the Royal Astronomical Society} \textbf{464}, 3409 (2017), arXiv: \eprint{1607.03149}.

\bibitem{sanchez2016clustering}
A.~G. {S{\'a}nchez}, J.~N. {Grieb}, S.~{Salazar-Albornoz}, S.~{Alam}, F.~{Beutler}, A.~J. {Ross}, J.~R. {Brownstein}, C.-H. {Chuang}, A.~J. {Cuesta}, D.~J. {Eisenstein}, et~al., \href{http://dx.doi.org/10.1093/mnras/stw2495}{Monthly Notices of the Royal Astronomical Society} \textbf{464}, 1493 (2017), arXiv: \eprint{1607.03146}.

\bibitem{alam2017clustering}
S.~{Alam}, M.~{Ata}, S.~{Bailey}, F.~{Beutler}, D.~{Bizyaev}, J.~A. {Blazek}, A.~S. {Bolton}, J.~R. {Brownstein}, A.~{Burden}, C.-H. {Chuang}, et~al., \href{http://dx.doi.org/10.1093/mnras/stx721}{Monthly Notices of the Royal Astronomical Society} \textbf{470}, 2617 (2017), arXiv: \eprint{1607.03155}.

\bibitem{chuang2017clustering}
C.-H. {Chuang}, F.-S. {Kitaura}, Y.~{Liang}, A.~{Font-Ribera}, C.~{Zhao}, P.~{McDonald}, and C.~{Tao}, \href{http://dx.doi.org/10.1103/PhysRevD.95.063528}{Physical Review D} \textbf{95}, 063528 (2017), arXiv: \eprint{1605.05352}.

\bibitem{DESI}
{DESI Collaboration}, A.~{Aghamousa}, J.~{Aguilar}, S.~{Ahlen}, S.~{Alam}, L.~E. {Allen}, C.~{Allende Prieto}, J.~{Annis}, S.~{Bailey}, and C.~{Balland}, \href{https://ui.adsabs.harvard.edu/abs/2016arXiv161100036D}{arXiv e-prints} arXiv:1611.00036 (2016), arXiv: \eprint{1611.00036}.

\bibitem{Sabiu2016A&A}
C.~G. {Sabiu}, D.~F. {Mota}, C.~{Llinares}, and C.~{Park}, \href{http://dx.doi.org/10.1051/0004-6361/201527776}{Astronomy and Astrophysics} \textbf{592}, A38 (2016), arXiv: \eprint{1603.05750}.

\bibitem{Slepian_2017}
Z.~{Slepian}, D.~J. {Eisenstein}, J.~R. {Brownstein}, C.-H. {Chuang}, H.~{Gil-Mar{\'\i}n}, S.~{Ho}, F.-S. {Kitaura}, W.~J. {Percival}, A.~J. {Ross}, G.~{Rossi}, et~al., \href{http://dx.doi.org/10.1093/mnras/stx488}{Monthly Notices of the Royal Astronomical Society} \textbf{469}, 1738 (2017), arXiv: \eprint{1607.06097}.

\bibitem{Sabiu_2019}
C.~G. {Sabiu}, B.~{Hoyle}, J.~{Kim}, and X.-D. {Li}, \href{http://dx.doi.org/10.3847/1538-4365/ab22b5}{The Astrophysical Journals} \textbf{242}, 29 (2019), arXiv: \eprint{1901.00296}.

\bibitem{ryden1995measuring}
B.~S. {Ryden}, \href{http://dx.doi.org/10.1086/176277}{The Astrophysical Journal} \textbf{452}, 25 (1995), arXiv: \eprint{astro-ph/9506028}.

\bibitem{lavaux2012precision}
G.~{Lavaux} and B.~D. {Wandelt}, \href{http://dx.doi.org/10.1088/0004-637X/754/2/109}{The Astrophysical Journal} \textbf{754}, 109 (2012), arXiv: \eprint{1110.0345}.

\bibitem{Ravanbakhsh17}
S.~{Ravanbakhsh}, J.~{Oliva}, S.~{Fromenteau}, L.~C. {Price}, S.~{Ho}, J.~{Schneider}, and B.~{Poczos}, \href{https://ui.adsabs.harvard.edu/abs/2017arXiv171102033R}{arXiv e-prints}  (2017), arXiv: \eprint{1711.02033}.

\bibitem{Mathuriya18}
A.~{Mathuriya}, D.~{Bard}, P.~{Mendygral}, L.~{Meadows}, J.~{Arnemann}, L.~{Shao}, S.~{He}, T.~{Karna}, D.~{Moise}, S.~J. {Pennycook}, et~al., \href{https://ui.adsabs.harvard.edu/abs/2018arXiv180804728M}{arXiv e-prints}  (2018), arXiv: \eprint{1808.04728}.

\bibitem{pan2020cosmological}
S.~Pan, M.~Liu, J.~Forero-Romero, C.~G. Sabiu, Z.~Li, H.~Miao, and X.-D. Li, Science China Physics, Mechanics \& Astronomy \textbf{63}, 110412 (2020).

\bibitem{wang2023darkai}
Z.~Wang, F.~Shi, X.~Yang, Q.~Li, Y.~Liu, and X.~Li, arXiv preprint arXiv:2305.11431  (2023).

\bibitem{Beisbart:2000ja}
C.~{Beisbart} and M.~{Kerscher}, \href{http://dx.doi.org/10.1086/317788}{The Astrophysical Journal} \textbf{545}, 6 (2000), arXiv: \eprint{astro-ph/0003358}.

\bibitem{Beisbart2002}
C.~{Beisbart}, M.~{Kerscher}, and K.~{Mecke}, \emph{{Mark Correlations: Relating Physical Properties to Spatial Distributions}}, volume 600, 358--390 (2002).

\bibitem{Gottl2002}
S.~{Gottl{\"o}ber}, M.~{Kerscher}, A.~V. {Kravtsov}, A.~{Faltenbacher}, A.~{Klypin}, and V.~{M{\"u}ller}, \href{http://dx.doi.org/10.1051/0004-6361:20020339}{Astronomy and Astrophysics} \textbf{387}, 778 (2002), arXiv: \eprint{astro-ph/0203148}.

\bibitem{Sheth:2004vb}
R.~K. {Sheth} and G.~{Tormen}, \href{http://dx.doi.org/10.1111/j.1365-2966.2004.07733.x}{Monthly Notices of the Royal Astronomical Society} \textbf{350}, 1385 (2004), arXiv: \eprint{astro-ph/0402237}.

\bibitem{Sheth:2005aj}
R.~K. {Sheth}, A.~J. {Connolly}, and R.~{Skibba}  (2005), arXiv: \eprint{astro-ph/0511773}.

\bibitem{Skibba2006}
R.~{Skibba}, R.~K. {Sheth}, A.~J. {Connolly}, and R.~{Scranton}, \href{http://dx.doi.org/10.1111/j.1365-2966.2006.10196.x}{Monthly Notices of the Royal Astronomical Society} \textbf{369}, 68 (2006), arXiv: \eprint{astro-ph/0512463}.

\bibitem{White_2009}
M.~{White} and N.~{Padmanabhan}, \href{http://dx.doi.org/10.1111/j.1365-2966.2009.14732.x}{Monthly Notices of the Royal Astronomical Society} \textbf{395}, 2381 (2009), arXiv: \eprint{0812.4288}.

\bibitem{White2016}
M.~{White}, \href{http://dx.doi.org/10.1088/1475-7516/2016/11/057}{JCAP} \textbf{2016}, 057 (2016), arXiv: \eprint{1609.08632}.

\bibitem{Satpathy:2019nvo}
S.~{Satpathy}, R.~{A C Croft}, S.~{Ho}, and B.~{Li}, \href{http://dx.doi.org/10.1093/mnras/stz009}{Monthly Notices of the Royal Astronomical Society} \textbf{484}, 2148 (2019), arXiv: \eprint{1901.01447}.

\bibitem{massara2020}
E.~Massara, F.~Villaescusa-Navarro, S.~Ho, N.~Dalal, and D.~N. Spergel, Physical Review Letters \textbf{126}, 011301 (2021).

\bibitem{Philcox2020}
O.~H.~E. Philcox, E.~Massara, and D.~N. Spergel  (2020), arXiv: \eprint{2006.10055}.

\bibitem{MCF_Yang}
Y.~Yang, H.~Miao, Q.~Ma, M.~Liu, C.~G. Sabiu, J.~Forero-Romero, Y.~Huang, L.~Lai, Q.~Qian, Y.~Zheng, et~al., The Astrophysical Journal \textbf{900}, 6 (2020).

\bibitem{SDSS_Reid}
B.~{Reid}, S.~{Ho}, N.~{Padmanabhan}, W.~J. {Percival}, J.~{Tinker}, R.~{Tojeiro}, M.~{White}, D.~J. {Eisenstein}, C.~{Maraston}, A.~J. {Ross}, et~al., \href{http://dx.doi.org/10.1093/mnras/stv2382}{Monthly Notices of the Royal Astronomical Society} \textbf{455}, 1553 (2016), arXiv: \eprint{1509.06529}.

\bibitem{rodriguez2016clustering_bigmd}
S.~A. Rodr{\'\i}guez-Torres, C.-H. Chuang, F.~Prada, H.~Guo, A.~Klypin, P.~Behroozi, C.~H. Hahn, J.~Comparat, G.~Yepes, A.~D. Montero-Dorta, et~al., Monthly Notices of the Royal Astronomical Society \textbf{460}, 1173 (2016).

\bibitem{kim2015horizon}
J.~{Kim}, C.~{Park}, B.~{L'Huillier}, and S.~E. {Hong}, \href{http://dx.doi.org/10.5303/JKAS.2015.48.4.213}{Journal of Korean Astronomical Society} \textbf{48}, 213 (2015), arXiv: \eprint{1508.05107}.

\bibitem{Jiang_2008_J08}
C.~Y. {Jiang}, Y.~P. {Jing}, A.~{Faltenbacher}, W.~P. {Lin}, and C.~{Li}, \href{http://dx.doi.org/10.1086/526412}{The Astrophysical Journal} \textbf{675}, 1095 (2008), arXiv: \eprint{0707.2628}.

\bibitem{Li16}
X.-D. {Li}, C.~{Park}, C.~G. {Sabiu}, H.~{Park}, D.~H. {Weinberg}, D.~P. {Schneider}, J.~{Kim}, and S.~E. {Hong}, \href{http://dx.doi.org/10.3847/0004-637X/832/2/103}{The Astrophysical Journal} \textbf{832}, 103 (2016), arXiv: \eprint{1609.05476}.

\bibitem{tassev2013_COLA}
S.~Tassev, M.~Zaldarriaga, and D.~J. Eisenstein, \href{http://dx.doi.org/10.1088/1475-7516/2013/06/036}{Journal of Cosmology and Astroparticle Physics} \textbf{2013}, 036–036 (2013), \urlprefix\url{http://dx.doi.org/10.1088/1475-7516/2013/06/036}.

\bibitem{kitaura2016clustering_Patchy}
F.-S. Kitaura, S.~Rodriguez-Torres, C.-H. Chuang, C.~Zhao, F.~Prada, H.~Gil-Marin, H.~Guo, G.~Yepes, A.~Klypin, C.~G. Sc{\'o}ccola, et~al., Monthly Notices of the Royal Astronomical Society \textbf{456}, 4156 (2016).

\bibitem{behroozi2012rockstar}
P.~S. {Behroozi}, R.~H. {Wechsler}, and H.-Y. {Wu}, \href{http://dx.doi.org/10.1088/0004-637X/762/2/109}{The Astrophysical Journal} \textbf{762}, 109 (2013), arXiv: \eprint{1110.4372}.

\bibitem{gunn1998sloan}
J.~E. Gunn, M.~Carr, C.~Rockosi, M.~Sekiguchi, K.~Berry, B.~Elms, E.~De~Haas, {\v{Z}}.~Ivezi{\'c}, G.~Knapp, R.~Lupton, et~al., The Astronomical Journal \textbf{116}, 3040 (1998).

\bibitem{fukugita1996sloan}
M.~Fukugita, K.~Shimasaku, T.~Ichikawa, J.~Gunn, et~al., The sloan digital sky survey photometric system, Technical report, SCAN-9601313 (1996).

\bibitem{eisenstein2011sdss}
D.~J. Eisenstein, D.~H. Weinberg, E.~Agol, H.~Aihara, C.~A. Prieto, S.~F. Anderson, J.~A. Arns, {\'E}.~Aubourg, S.~Bailey, E.~Balbinot, et~al., The Astronomical Journal \textbf{142}, 72 (2011).

\bibitem{dawson2012baryon_BOSS}
K.~S. Dawson, D.~J. Schlegel, C.~P. Ahn, S.~F. Anderson, {\'E}.~Aubourg, S.~Bailey, R.~H. Barkhouser, J.~E. Bautista, A.~Beifiori, A.~A. Berlind, et~al., The Astronomical Journal \textbf{145}, 10 (2012).

\bibitem{smee2013multi_BOSS}
S.~A. Smee, J.~E. Gunn, A.~Uomoto, N.~Roe, D.~Schlegel, C.~M. Rockosi, M.~A. Carr, F.~Leger, K.~S. Dawson, M.~D. Olmstead, et~al., The Astronomical Journal \textbf{146}, 32 (2013).

\bibitem{LI14}
X.-D. {Li}, C.~{Park}, J.~E. {Forero-Romero}, and J.~{Kim}, \href{http://dx.doi.org/10.1088/0004-637X/796/2/137}{The Astrophysical Journal} \textbf{796}, 137 (2014), arXiv: \eprint{1412.3564}.

\bibitem{LI15}
X.-D. {Li}, C.~{Park}, C.~G. {Sabiu}, and J.~{Kim}, \href{http://dx.doi.org/10.1093/mnras/stv622}{Monthly Notices of the Royal Astronomical Society} \textbf{450}, 807 (2015), arXiv: \eprint{1504.00740}.

\bibitem{LI18}
X.-D. {Li}, C.~G. {Sabiu}, C.~{Park}, Y.~{Wang}, G.-b. {Zhao}, H.~{Park}, A.~{Shafieloo}, J.~{Kim}, and S.~E. {Hong}, \href{http://dx.doi.org/10.3847/1538-4357/aab42e}{The Astrophysical Journal} \textbf{856}, 88 (2018), arXiv: \eprint{1803.01851}.

\bibitem{LI19}
X.-D. {Li}, H.~{Miao}, X.~{Wang}, X.~{Zhang}, F.~{Fang}, X.~{Luo}, Q.-G. {Huang}, and M.~{Li}, \href{http://dx.doi.org/10.3847/1538-4357/ab0f30}{The Astrophysical Journal} \textbf{875}, 92 (2019), arXiv: \eprint{1903.04757}.

\bibitem{Park:2019mvn}
H.~{Park}, C.~{Park}, C.~G. {Sabiu}, X.-d. {Li}, S.~E. {Hong}, J.~{Kim}, M.~{Tonegawa}, and Y.~{Zheng}, \href{http://dx.doi.org/10.3847/1538-4357/ab2da1}{The Astrophysical Journal} \textbf{881}, 146 (2019), arXiv: \eprint{1904.05503}.

\bibitem{Zhang2019}
Z.~{Zhang}, G.~{Gu}, X.~{Wang}, Y.-H. {Li}, C.~G. {Sabiu}, H.~{Park}, H.~{Miao}, X.~{Luo}, F.~{Fang}, and X.-D. {Li}, \href{http://dx.doi.org/10.3847/1538-4357/ab1ea4}{The Astrophysical Journal} \textbf{878}, 137 (2019), arXiv: \eprint{1902.09794}.

\bibitem{collaboration2014planck}
P.~Collaboration, P.~Ade, N.~Aghanim, C.~Armitage-Caplan, M.~Arnaud, M.~Ashdown, F.~Atrio-Barandela, J.~Aumont, C.~Baccigalupi, A.~Banday, et~al., A\&A \textbf{571}, A16 (2014).

\bibitem{Berlind_2002_HOD}
A.~A. {Berlind} and D.~H. {Weinberg}, \href{http://dx.doi.org/10.1086/341469}{The Astrophysical Journal} \textbf{575}, 587 (2002), arXiv: \eprint{astro-ph/0109001}.

\bibitem{klypin2016multidark_patchy}
A.~Klypin, G.~Yepes, S.~Gottl{\"o}ber, F.~Prada, and S.~Hess, Monthly Notices of the Royal Astronomical Society \textbf{457}, 4340 (2016).

\bibitem{MCF_Xiao}
X.~Xiao, Y.~Yang, X.~Luo, J.~Ding, Z.~Huang, X.~Wang, Y.~Zheng, C.~G. Sabiu, J.~Forero-Romero, H.~Miao, et~al., \href{http://dx.doi.org/10.1093/mnras/stac879}{Monthly Notices of the Royal Astronomical Society} \textbf{513}, 595 (2022), \urlprefix\url{https://academic.oup.com/mnras/article-abstract/513/1/595/6562085}.

\bibitem{Cai_2010_MCF}
Y.-C. Cai, G.~Bernstein, and R.~K. Sheth, \href{http://dx.doi.org/10.1111/j.1365-2966.2010.17969.x}{Monthly Notices of the Royal Astronomical Society} no--no (2010), \urlprefix\url{https://doi.org/10.1111/j.1365-2966.2010.17969.x}.

\bibitem{Chan_2022}
K.~C. Chan, I.~Ferrero, S.~Avila, A.~J. Ross, M.~Crocce, and E.~Gazta{\~{n} }aga, \href{http://dx.doi.org/10.1093/mnras/stac340}{Monthly Notices of the Royal Astronomical Society} \textbf{511}, 3965 (2022), \urlprefix\url{https://academic.oup.com/mnras/article/511/3/3965/6526328}.

\bibitem{Gingold1977}
R.~A. {Gingold} and J.~J. {Monaghan}, \href{http://dx.doi.org/10.1093/mnras/181.3.375}{Monthly Notices of the Royal Astronomical Society} \textbf{181}, 375 (1977).

\bibitem{Lucy1977}
L.~B. {Lucy}, \href{http://dx.doi.org/10.1086/112164}{The Astronomical Journal} \textbf{82}, 1013 (1977).

\bibitem{Landy}
S.~D. {Landy} and A.~S. {Szalay}, \href{http://dx.doi.org/10.1086/172900}{The Astrophysical Journal} \textbf{412}, 64 (1993).

\bibitem{Petri_2016_PCA}
A.~Petri, M.~May, and Z.~Haiman, \href{http://dx.doi.org/10.1103/physrevd.94.063534}{Physical Review D} \textbf{94} (2016), \urlprefix\url{https://doi.org/10.1103%2Fphysrevd.94.063534}.

\bibitem{Planck2020}
{Planck Collaboration}, N.~{Aghanim}, Y.~{Akrami}, M.~{Ashdown}, J.~{Aumont}, C.~{Baccigalupi}, M.~{Ballardini}, A.~J. {Banday}, R.~B. {Barreiro}, N.~{Bartolo}, et~al., \href{http://dx.doi.org/10.1051/0004-6361/201833910}{Astronomy and Astrophysics} \textbf{641}, A6 (2020), arXiv: \eprint{1807.06209}.

\end{thebibliography}
\end{multicols}
\vspace*{-4mm}
\end{document}